\documentclass[aps,reprint,superscriptaddress,eqsecnum,nofootinbib,amsmath,amssymb,prd,floatfix]{revtex4-2}

\usepackage[utf8]{inputenc}
\usepackage[T1]{fontenc}
\usepackage[dvipsnames]{xcolor}
\usepackage{graphicx}
\usepackage[colorlinks=true,allcolors=teal]{hyperref}
\usepackage[normalem]{ulem}
\usepackage{orcidlink}
\usepackage{units}
\usepackage{multirow}

\newcommand{\ud}{\mathrm{d}}

\newcommand{\DM}{\mathrm{DM}}

\newcommand{\GW}{\mathrm{GW}}
\newcommand{\I}{\mathrm{i}}
\newcommand{\IN}{\mathrm{in}}
\newcommand{\MIN}{\mathrm{min}}
\newcommand{\MAX}{\mathrm{max}}
\newcommand{\ISCO}{\mathrm{ISCO}}

\newcommand{\SP}{\mathrm{sp}}

\newcommand{\fy}{\mathrm{2,4y}}

\newcommand{\Msolar}{\mathrm M_\odot}
\newcommand{\Jmin}{\mathcal J_\MIN}

\newcommand{\UVA}{Department of Physics, University of Virginia, P.O.~Box 400714, Charlottesville, Virginia 22904-7414, USA}

\begin{document}

\title{Effects of the interaction of dark matter and neutron-star matter on extreme and intermediate mass-ratio inspirals}

\author{Benjamin A.\ Wade\,\orcidlink{0009-0000-2764-5165}}
\email{baw8td@virginia.edu}
\affiliation{\UVA}%

\author{Julian Heeck\,\orcidlink{0000-0003-2653-5962}}
\email{heeck@virginia.edu}
\affiliation{\UVA}%

\author{David A.\ Nichols\,\orcidlink{0000-0002-4758-9460}}
\email{david.nichols@virginia.edu}
\affiliation{\UVA}%

\date{\today}

\begin{abstract}
Extreme and intermediate mass-ratio inspirals in a dense dark-matter distribution have the effects of the dark matter imprinted on the orbital dynamics of and the emitted gravitational waves from these systems. Prior work has shown that space-based gravitational-wave detectors can measure the dark-matter-induced effects on the gravitational waves, which would give evidence for the presence of dark matter around the massive black hole. In this earlier work, the dark matter has been assumed to have only gravitational interactions (namely, no dark-matter self-annihilation or interactions between dark matter and ordinary baryonic or leptonic matter). In this article, we investigate the gravitational-wave effects of introducing such interactions of dark matter with itself or with ordinary matter in binaries with a neutron-star secondary. We consider broad classes of dark-matter models that change the distribution of dark matter (spikes or annihilation plateaus) and which permit accretion onto the secondary, as well as an increasing or static mass of the neutron star during the inspiral (in addition to the purely gravitational effect of dynamical friction). We find distinctive gravitational-wave signatures of these interactions and of self-annihilation, which in some of the scenarios could be sufficiently large for space-based detectors to distinguish them using gravitational-wave observations of these systems.
\end{abstract}

\maketitle

\tableofcontents

\section{Introduction} \label{sec:intro}

The LIGO-Virgo-KAGRA Collaboration has now announced the observation of gravitational waves (GWs) from the mergers of around three hundred compact binaries~\cite{LIGOScientific:2018mvr,LIGOScientific:2020ibl,KAGRA:2021vkt,LIGOScientific:2025slb,LIGOScientific:2026wfs}.
The majority of these events have been produced by the mergers of two stellar-mass black holes (BHs), but a handful are consistent with BH-neutron star (NS) mergers~\cite{LIGOScientific:2021qlt} and NS-NS mergers~\cite{LIGOScientific:2017vwq}.
Collectively, these GW events have given a first view and a more detailed understanding of the population of the known astrophysical GW sources at the high-frequency end of the GW spectrum.
At the lower end of this spectrum, pulsar timing arrays have found strong evidence in favor of a stochastic background of gravitational waves arising from the mergers of supermassive BHs (which reside in the centers of most galaxies)~\cite{NANOGrav:2023gor,EPTA:2023fyk,Reardon:2023gzh,Xu:2023wog}.
However, there is a large gap in the GW spectrum between the upper limit of pulsar timing arrays' spectral sensitivity ($\sim \unit[10^{-6}]{Hz}$) and the lower limit of that of the LIGO-Virgo-KAGRA detectors ($\sim \unit[10]{Hz}$).

A significant portion of this gap in the GW spectrum in the $\sim \unit[10^{-4}\text{--}10^{-1}]{Hz}$ range will be filled by the Laser Interferometer Space Antenna (LISA) mission~\cite{Amaro-Seoane:2017ADS,Baker:2019nia}.
In addition to mergers of supermassive BH binaries, LISA is expected to measure galactic white-dwarf binaries and extragalactic mergers of stellar-mass compact objects with supermassive and intermediate-mass BHs.
Members of this last class of system are called extreme mass-ratio inspirals (EMRIs), or intermediate mass-ratio inspirals (IMRIs) if rather than a supermassive BH, the more massive object in the binary is an intermediate-mass BH (IMBH)~\cite{LISA:2022yao}.
When EMRIs merge in vacuum, they can undergo $O(10^5)$ orbits during a 4-year LISA observation window as the secondary inspirals toward the massive BH in the most strongly curved region of the spacetime outside of the BH's event horizon.
The GWs emitted from the EMRI contain sufficient information about this strongly curved spacetime geometry to allow a GW measurement to determine whether the massive object is consistent with a BH solution in general relativity to high accuracy~\cite{Ryan:1995wh} (which is a result referred to as ``Ryan's theorem'' in the literature~\cite{Li:2007qu}).
Conversely, if the EMRI merger takes place in an astrophysical environment containing (for example) baryonic matter, dark matter (DM), or third bodies producing tidal effects, the effects of its environs can be imprinted on the EMRIs dynamics and the emitted GWs (if the density of matter is sufficiently high or the tidal effects are sufficiently strong~\cite{Barausse:2014tra,Cardoso:2019rou,Cardoso:2022whc,Figueiredo:2023gas}).

This paper will focus on the effects of dense DM environments on the GWs emitted from IMRI systems, which is a topic that has been investigated in some detail (\cite{Eda:2013gg,Eda:2014kra,Yue:2017iwc,Edwards:2019tzf,Kavanagh:2020cfn,Coogan:2021uqv,Becker:2021ivq,Speeney:2022ryg,Cole:2022yzw,Nichols:2023ufs,Berezhiani:2023vlo,Mukherjee:2023lzn,Wilcox:2024sqs,Speeney:2024mas,Bertone:2024wbn,Karydas:2024fcn,Kavanagh:2024lgq,Wade:2025rkk,Karydas:2025gqp,Mitra:2025tag}, for example). 
For the DM to have a measurable effect on the emitted GWs, its density near the massive BH needs to be higher than the densities in the centers of both cored and cuspy DM halos, such as the Navarro--Frenk--White profile~\cite{Navarro:1995iw} in the latter case.
These higher densities can form through adiabatic growth of a seed BH in a DM halo, in which case the DM profile is referred to as a ``DM spike''~\cite{Gondolo:1999ef,Sadeghian:2013laa,Ferrer:2017xwm}.
An alternate mechanism for reaching such high densities was proposed in~\cite{Bertone:2024wbn} and was referred to as a ``DM mound.''
The DM spikes have been discussed extensively in the literature, both in the contexts of amplified indirect detection signatures from DM annihilation in the spike and of the impact on the GWs in IMRI and EMRI systems (see, e.g.,~\cite{Bertone:2024rxe} for a review).

The survival of DM spikes under a variety of different astrophysical processes has been discussed in the literature, and there are many astrophysical processes that can deplete the density in these spikes (see~\cite{Ullio:2001fb}).
These mechanisms were more likely to apply for more massive BHs, which is why more recent work focused on IMBHs and IMRI systems where the DM spike or mound has a better chance to survive.
More recently, it was investigated how stellar populations in nuclear clusters and past IMRI and EMRI systems around the IMBH can affect a DM spike~\cite{Sharpe:2026nqq,Karydas:2026ctq}.
There are calculations suggesting that stellar populations with more realistic distributions of stellar masses will deplete the spike density at radii much larger than those typical of EMRI systems that LISA will measure.
Whether EMRIs or IMRIs deplete the density in the region of spacetime that is relevant for the emitted GWs that LISA will measure is not yet determined~\cite{Sharpe:2026nqq,Karydas:2026ctq}.
We work under the assumption that the density is not significantly depleted in this paper.

The question of whether a large number of IMRI and EMRI events over cosmic time can significantly modify the DM spike is related to the relevant physical effects that should be included while modeling the binary in a DM environment.
The early work~\cite{Eda:2013gg,Eda:2014kra,Yue:2017iwc} evolved just the IMRI, not the surrounding DM, and this neglect of feedback to the DM spike from \emph{dynamical friction} (DF)~\cite{Chandrasekhar:1943ys} led to significant overestimates of the DM effects on the IMRI's orbit~\cite{Kavanagh:2020cfn}.
Reference~\cite{Kavanagh:2020cfn} assumed the secondary in the IMRI was a NS, but it assumed just gravitational interactions between the DM particle and the NS, and it neglected the strong scattering of DM particles through the NS interior (which~\cite{Karydas:2025gqp} showed can cause a substantial difference).
For BH secondaries, accretion of DM particles onto the secondary and the associated feedback on the DM distribution were shown to be important to include in~\cite{Nichols:2023ufs,Karydas:2024fcn}.
We will refer to this effect as \emph{secondary accretion} (SA), as in~\cite{Nichols:2023ufs,Wade:2025rkk}, to avoid confusion with accretion onto the primary (which is expected to have negligible effects on the IMRI's orbit during the year-long timescales during which LISA will measure the system~\cite{Nichols:2023ufs}).
For either NS or BH secondaries, there have been recent investigations into the contribution of three-body slingshot effects~\cite{Mukherjee:2023lzn} and time-dependent asphericity in the IMRI's potential (called ``stirring'' in~\cite{Kavanagh:2024lgq}) on the DM distribution, and what changes it induces from the procedure used in~\cite{Kavanagh:2020cfn,Nichols:2023ufs,Karydas:2024fcn,Wade:2025rkk}.\footnote{The uncertainties about the effects of multiple EMRI mergers on the long-term fate of the DM spike in~\cite{Sharpe:2026nqq,Karydas:2026ctq} hinge upon whether ejection of DM particles by the slingshot effects or redistribution of DM due to feedback from dynamical friction and the stirring effect are the more important dynamical effects during the late-time evolution of the IMRI or EMRI.}

In this paper, we follow the general approach and formalism used in~\cite{Nichols:2023ufs,Wade:2025rkk} for BH secondaries (which includes DF and SA, but does not model slingshot effects or stirring), but we adapt the formalism so as to study IMRIs with a NS secondary.
We consider different classes of DM models that have DM-matter interactions and DM-DM annihilation both in the DM spike and in DM particles that are captured by the NS during the inspiral.
We focus on these effects on the IMRI's orbit, the GW phase, and the post-merger DM density.
We summarize the organization of our paper and main results next.

\subsection{Organization and results of this paper}

In Sec.~\ref{sec:DMmodels_spike}, we review DM models where there is no DM-DM annihilation in the DM spike, so that it initially has the same form as the spike with the angular momentum cutoff given in~\cite{Wade:2025rkk} (for the first-generation mergers).
We consider three scenarios: one in which there are no DM-NS matter interactions (similar to that in~\cite{Kavanagh:2020cfn}), a second where DM-NS matter interactions cause the DM to be efficiently captured in the NS (asymmetric DM), and a third where the DM is captured, but DM-NS matter interactions mediate the annihilation of DM particles into other particles that can escape the NS (and consequently the mass of the NS does not increase).
In Sec.~\ref{sec:DMmodels_plateau}, we consider three similar scenarios, but now in DM models with DM-DM annihilation, which causes an annihilation plateau in the initial DM distribution.
In Sec.~\ref{sec:evolution}, we review the joint evolution equations for the IMRI's orbit and the DM distribution function.
We discuss the differences that arise from the methods used in~\cite{Wade:2025rkk} because of DM-DM annihilation and DM-NS matter interactions.
Section~\ref{sec:methods} discusses our numerical methods and details of our simulations.
Our main results for the GW phase from these systems and the dynamics of the DM are given in Sec.~\ref{sec:results}.
Our conclusions and some brief discussion are presented in Sec.~\ref{sec:conclusions}, though we give a brief overview of some of our findings next.

The different cases of DM-NS matter interactions produce distinctive changes in the dynamics of the IMRI and the corresponding GWs emitted from these systems.
In DM spikes, no effects of SA are present with only gravitational interactions, but adding interactions adds the effects of SA (and the corresponding mass increase in one case) on the orbital evolution and GW phase.
In DM plateaus, these similar effects arise in the three scenarios of DM-NS matter interactions, but the annihilation plateau can significantly deplete the DM particles that move more slowly than the orbital speed of the secondary.
This, in turn, can greatly diminish the effects of dynamical friction, and leads to scenarios in which the GW dephasing is generated by SA accretion and the corresponding mass increase.
These effects on the GW phase can be sufficiently large for LISA to have the potential to distinguish several of these six scenarios.

\section{Dark-matter models that permit a spike density profile} \label{sec:DMmodels_spike}

In this section and Sec.~\ref{sec:DMmodels_plateau}, we assume cold particle-like DM in the typical mass range for weakly-interacting massive particles, say from roughly one MeV/c$^2$ to tens of TeV/c$^2$~\cite{Cirelli:2024ssz}. 
This restriction on the mass range is not crucial for the strictly gravitational phenomena, such as dynamical friction; it is more relevant for the DM phenomenology described below. 
Rather than focusing on one particular model, we will sketch different \emph{classes} of models that lead to qualitatively different effects on the IMRI's orbital dynamics.
In all cases, we assume that the primary BH has mass $m_1$, is nonrotating, and is located at the center of a spherically-symmetric DM distribution.

For the DM spike density profiles in this section, we take the initial density to be a function of radius $r$ from the primary BH of the form
\begin{align} \label{eq:rhoDM}
    \rho_\DM(r)=
\begin{cases}
\rho_{\rm sp}\left(\dfrac{r_{\rm sp}}{r}\right)^{\gamma_{\rm sp}} \left(1-\dfrac{r_\text{in}}{r}\right)^{\gamma_{\rm sp}},
& r_{\rm in} \le r \le r_{\rm sp} \,,\\
0\,,
& r < r_{\rm in}\, .
\end{cases}
\end{align}
The exact values of the DM spike's power law exponent $\gamma_\SP$ and the normalization of the density $\rho_\SP$ depend on the BH's formation history and environment (see~\cite{Gondolo:1999ef}).
We will use the values from~\cite{Eda:2014kra}, which assumed a small seed BH that grew in mass adiabatically in the center of a Navarro--Frenk--White profile~\cite{Navarro:1995iw}.
This scenario produces $\rho_\SP \approx \unit[200]{\Msolar/pc^3}$ and $\gamma_\SP \approx 7/3$.
The inner radius $r_\text{in} \equiv 4 G m_1/c^2$ exists because DM particles that are gravitationally bound to $m_1$ and that pass within this radius would be accreted onto the primary BH~\cite{Sadeghian:2013laa} (because the particle's angular momentum is too low).
This is reviewed in more detail in Sec.~\ref{subsec:DMdensity}.

The term $(1-r_\IN/r)$ in Eq.~\eqref{eq:rhoDM} implements an appropriate angular-momentum cutoff for the DM particles in Newtonian gravity~\cite{Wade:2025rkk} and serves as a simple approximation to the fully relativistic result of~\cite{Sadeghian:2013laa}. 
The outer radius of the spike, $r_\SP$, will be chosen as in~\cite{Eda:2014kra}: namely, one fifth the radius at which the total enclosed DM mass equals twice the mass $m_1$,
\begin{align} \label{eq:r_sp}
    r_{\rm sp} \approx
\left[
\frac{0.2^{3-\gamma_{\rm sp}}(3-\gamma_{\rm sp})m_1}
{2\pi \rho_{\rm sp}}
\right]^{1/3} .
\end{align}
For $\rho_\DM(r)$ at $r > r_\text{sp}$, the DM density smoothly transitions to the inner portion of a typical galactic DM distribution, in this case a Navarro--Frenk--White profile with $\rho_\DM\propto r^{-1}$. 
This outer region will not be needed to analyze the IMRI and EMRI systems in this paper.

In the remainder of this section, we will sketch DM models that are consistent with such a spike profile around the primary BH. 
Importantly, these DM particles cannot annihilate outside of the NS; otherwise, the high DM spike densities would be decreased.
We leave discussion of such DM models with DM-DM annihilation to Sec.~\ref{sec:DMmodels_plateau}.

\subsection{Dark matter with purely gravitational interactions}
\label{sec:spike}

The current observational evidence for DM is compatible with DM particles that interact via the gravitational force only (i.e., with no, or at least highly suppressed, DM-matter and DM-DM interactions~\cite{Cirelli:2024ssz}). 
Such a scenario severely limits the capability of direct DM detection experiments to identify the quantum numbers of DM (e.g., its mass and spin).
However, models of this kind would still allow for DM-induced effects on the GWs emitted from EMRI systems (as reviewed in Sec.~\ref{sec:intro} and described in more detail in Sec.~\ref{sec:evolution}).

In particular, this scenario consists of a supermassive BH which adiabatically grows in a DM halo to form a DM spike, which can achieve high DM densities because there is no DM annihilation to flatten the DM density.
A secondary compact object (a BH or NS) moving through the DM halo experiences DF~\cite{Chandrasekhar:1943ys}, which shortens the inspiral compared to mergers in vacuum. 
In addition, a BH secondary accretes via SA~\cite{Yue:2017iwc,Nichols:2023ufs,Karydas:2024fcn}, because some DM particles are on orbits near the secondary with a sufficiently small impact parameter to ultimately fall into the secondary BH's event horizon.
A low-velocity limit of the relativistic capture calculation shows that the SA cross section is given by
\begin{align}
    \sigma_\text{BH}(v_2) = \frac{16\pi (G m_2)^2}{(c v_2)^2} = 4\pi R_s^2\frac{c^2}{v_2^2}
    \label{eq:BH_cross_section}
\end{align}
(see~\cite[Exercise 25.22]{Misner:1973prb} or~\cite{Unruh:1976fm}).
The second equality in Eq.~\eqref{eq:BH_cross_section} was obtained by expressing the cross section in terms of the Schwarzschild radius $R_s=2G m_2/c^2$. 
The mass $m_2$ and velocity $v_2$ are those of the secondary BH, respectively; the latter of which was used as a proxy for the relative velocity of the DM particle and BH in~\cite{Yue:2017iwc,Nichols:2023ufs}.
For secondaries on a circular orbit, $v_2 \approx \sqrt{Gm_1/r_2}$, where $r_2$ is the magnitude of the orbital separation.
Secondary accretion leads to an increase in the secondary's mass given by
\begin{equation} \label{eq:dot_m2_BH}
    \dot{m}_2 = \rho_\text{DM}(r_2) \sigma_\text{BH}(v_2) v_2 ,
\end{equation}
and a faster inspiral due to angular-momentum conservation during accretion:
\begin{equation}
    \dot{r}_2^\text{SA} \approx -2 r_2 \dot{m}_2/m_2 
\end{equation}
(see \cite{Hughes:2018qxz,Nichols:2023ufs}). 
EMRIs with a BH secondary in a DM spike have been discussed extensively in the literature (e.g.,~\cite{Yue:2017iwc,Nichols:2023ufs,Wade:2025rkk}), and we will use this case primarily as a benchmark scenario for our comparisons with EMRIs with a NS secondary.

When the secondary is a NS, gravitational interactions are too weak to capture a sizable amount of DM in the NS, which effectively eliminates the mass increase due to SA (and the corresponding change in the orbital separation). 
The inspiral is driven by GW emission and the DM-induced changes to the inspiral from DF, which still is efficient.
Because there are DM particles which pass \emph{through} the NS, there are DM particles that gravitationally scatter with the interior solution of the NS (such strong-field scattering was neglected in the works~\cite{Eda:2014kra,Kavanagh:2020cfn,Coogan:2021uqv}, among others).
A more recent work,~\cite{Karydas:2025gqp}, computed the gravitational drag on the NS from such orbits.
Reference~\cite{Karydas:2025gqp} found that not only NS and BH secondaries had distinguishable inspirals,\footnote{\label{fn:BHmass}This is assuming that the BH and NS have the same mass so that they could not be distinguished by the secondary's mass; this also, of course, is a strong assumption given the observed distributions of NS and BH masses to date.} but NS with different equations of state did, too.
Thus, even in the purely gravitational case, there is the possibility to distinguish BH and NS secondaries from the stronger-field interactions of the DM particles with the spacetime of the secondary (subject to the caveat in Footnote~\ref{fn:BHmass}): specifically, SA for a BH has a distinctive signature from the DF arising in the NS interior.

In this paper, however, we will neglect the scattering in the interior, as was done in~\cite{Kavanagh:2020cfn}, so that any differences between a BH or a NS secondary will arise from the presence or absence of the effects of SA in the IMRI and DM distribution's evolution.
Adding DM-DM and DM-NS interactions, however, can further modify the DM-induced effects on the inspiral of an EMRI system by introducing some SA effects for a NS secondary.
We discuss these different DM scenarios and their effects in the next subsections.

\subsection{Asymmetric dark matter}
\label{sec:spike+SA}

Most theories of weakly-interacting massive particle DM include interactions of DM and ordinary-matter particles, usually as a means to explain the DM abundance in our universe~\cite{Cirelli:2024ssz}. 
With such interactions, there is the possibility of \textit{detecting} DM through, for example, DM-matter scattering (direct detection) or DM-DM annihilation into visible particles (indirect detection). 
Not observing these signatures in detectors puts stringent constraints on DM-matter or DM-DM interactions, but it cannot fully exclude these models.

In this subsection, we will focus on DM models that do not have (significant) DM-DM annihilation (e.g., asymmetric DM~\cite{Petraki:2013wwa}), so that a DM spike can be generated as discussed above; unlike the scenario in Sec.~\ref{sec:spike}, however, the DM here \textit{does} have interactions with matter inside the NS (e.g., DM couplings to neutrons, protons, electrons, or muons). 
These non-gravitational interactions allow the NS to capture DM.
To maximize the differences with the case of no SA for a NS (in Sec.~\ref{sec:spike}), we shall assume that these interactions are strong enough to capture \emph{all} DM in the path of the NS. 
Because of the extremely high matter density inside a NS, this does not even require a particularly large cross section.

As an order-of-magnitude estimate, the DM-neutron cross section just needs to be larger than $\pi R_\text{NS}^2/N_\text{NS}\simeq \pi (\unit[10]{km})^2/10^{57}\simeq \unit[3\times 10^{-45}]{cm^2}$, where $N_\text{NS}$ is the number of neutrons inside a typical NS and $R_\text{NS}$ is the NS radius. 
A more careful analysis changes this threshold cross section by at most an order of magnitude in the mass range $\unit{GeV}\lesssim m_\text{DM} \lesssim \unit[10^6]{GeV}$; however, lighter or heavier DM would require parametrically larger cross sections because scattering becomes more inefficient due to Pauli blocking or multiple scatterings becoming necessary~\cite{Anzuini:2021lnv}, respectively. 
Unlike in terrestrial experiments, the cross section can be velocity-dependent, inelastic, or spin-dependent and it would still allow for significant capture onto the NS since the DM particle accelerates to relativistic velocities as it falls into the NS's deep gravitational well~\cite{Bramante:2023djs}. 
Neutron stars also contain protons, electrons, and even muons~\cite{Garani:2019fpa} which have similar DM capture rates as neutrons~\cite{Bramante:2023djs}; this further disentangles capture in a NS from terrestrial direct-detection experiments. 
Thus, there exists a vast landscape of DM models that can evade direct-detection constraints, but which can also lead to significant DM capture in a NS.

Similarly to a BH, a NS will not just capture DM particles with impact parameters less than $R_\text{NS}$, but also larger impact parameters, which will get captured because of gravitational focusing~\cite{Kopp:2018jom}.
This enhances the \emph{capture rate} $C$ significantly above the geometric cross section.
For nonrelativistic secondaries, the rate is given by (see, e.g.,~\cite{Bell:2018pkk})
\begin{align}
    C \simeq \pi R_\text{NS}^2\frac{c^2}{v_2} \frac{v_\text{esc}^2/c^2}{1-v_\text{esc}^2/c^2} \frac{\rho_\text{DM}}{m_\text{DM}} \,,
    \label{eq:capture_rate}
\end{align}
where the NS escape velocity is $v_\text{esc} = \sqrt{2 G m_2/R_\text{NS}}$ and $v_2$ was again used as a proxy for the relative speed of the secondary and DM particle.
The NS's mass increase is given by
\begin{align}
    \dot{m}_2 = m_\text{DM} C \simeq \rho_\text{DM} \left( \pi R_\text{NS}^2\frac{c^2}{v_2^2} \frac{v_\text{esc}^2/c^2}{1-v_\text{esc}^2/c^2} \right) v_2 \, .
    \label{eq:accretion}
\end{align}
The expression has a similar qualitative form to Eq.~\eqref{eq:dot_m2_BH} for a BH, and it is thus natural to define a NS cross section by the quantity in the large parentheses
\begin{equation} \label{eq:sigma_NS}
    \sigma_\text{NS}(v_2) = \pi R_\text{NS}^2\frac{c^2}{v_2^2} \frac{v_\text{esc}^2/c^2}{1-v_\text{esc}^2/c^2} ,
\end{equation}
which can be compared to Eq.~\eqref{eq:BH_cross_section} for BHs. 
Both cross sections scale with the geometric cross section $\pi R^2$ (where $R$ is the surface of the NS or twice the radius of the event horizon in the Schwarzschild BH case),
and they are enhanced by $c^2/v_2^2$ due to gravitational focusing.
The additional factor in the NS case depends on the escape velocity, which will depend on $R_\text{NS}$ (which itself is a function of the NS equation of state).

As a simple estimate of the relative size of the BH and NS cross sections, we can use the Newtonian escape velocity $v_\mathrm{esc}^2 = 2Gm_2/R_\mathrm{NS}$.
Assuming that the BH and NS have the same mass $m_2$ then a comparison of Eqs.~\eqref{eq:BH_cross_section} and~\eqref{eq:sigma_NS} shows that the NS cross section is larger (i.e., $\sigma_\text{NS}\geq \sigma_\text{BH}$).
The equality occurs for $R_\mathrm{NS} = 4 Gm_2/c^2 = 2 R_s$.
For a NS of mass $2\Msolar$, for example, this corresponds to a NS radius of $\unit[12]{km}$. 

Realistic NSs with masses in the range $1.2 < m_2/M_\odot < 2.1$ have cross sections (and mass accretion rates) that are at most $10$--$20\%$ larger than a BH with the same mass, in this limit of efficient capture onto the NS.\footnote{One could also make the cross section $\sigma_\text{NS}$ arbitrarily small by reducing the DM-matter cross section below the threshold value mentioned above; this would restore the scenario considered in Sec.~\ref{sec:spike}.}
Since DF is similar for a BH and a NS with the same mass in this perfect-capture scenario, it will be challenging to distinguish BH--BH from BH--NS mergers via GWs when DM can be captured by the NS (more quantitative results will be given in Sec.~\ref{sec:results}).

In summary, there are viable DM models that can efficiently capture DM particles in NS and produce little to no DM annihilation that would disrupt the DM spike around the primary. 
The inspiral in this case would be dominated by GW emission, and both the effects of DF and SA on the inspiral would be similar to the case with a BH secondary.
Different phenomenology could arise if the captured DM accumulates in the NS core and changes the NS equation of state or even triggers a collapse into a BH (for bosonic DM)~\cite{Bramante:2023djs}. 
The high DM density in the spike could enhance these effects, but given that LISA will observe these systems for a timescale of order a few years, it is unlikely that the NS  would collapse to a BH during the final few years of the inspiral that LISA will measure.

\subsection{Annihilating dark matter in the neutron star}
\label{sec:spike+SA+no_growth}

The DM models of the previous section led to efficient capture on a NS and thus eventually a large DM density inside the NS (which is possible because of the high densities in the DM spike). 
Let us now assume that these captured DM particles can \textit{annihilate} inside the NS core into particles that escape the NS. 
Examples of annihilation products are low-energy neutrinos or other weakly-interacting particles. 
Since the accumulated DM component inside the NS is now evaporating away, the NS mass increase is suppressed compared to the usual capture scenarios. 
In an extreme limit, the NS does not accumulate mass at all, $\dot{m}_2 = 0$, as all captured DM is swiftly converted into isotropic radiation, e.g., in the form of neutrinos. 
Importantly, the NS still experiences a drag force from the one-sided DM collisions
(as will be discussed in Sec.~\ref{sec:evolution}, this produces a change in the orbital separation of the binary that goes as $\dot{r}_2 \simeq -2 r_2 \rho_\DM \sigma_\text{NS} v_2/m_2$). 

For most DM models, this would be an inconsistent setup because DM annihilations would also take place in the spike, significantly decrease the DM densities there, and leave fewer DM particles for the NS to capture. 
However, the DM annihilation rates in the spike and inside the NS core need not be the same: for example, the larger velocities in the DM spike compared to the NS core could change the rates of the velocity-dependent cross sections, or there could be scenarios in which DM annihilation depends on the surrounding neutron density, similar in spirit to models studied in Refs.~\cite{Boddy:2012xs,Davoudiasl:2022ubh,Davoudiasl:2026mqk}, which could enhance DM annihilation in the NS core while keeping the DM spike near the IMBH.

As a benchmark point for such, admittedly unusual, DM models, we assume IMRIs with a NS secondary in a DM spike, with DF, SA effects on $r_2$, but $\dot{m}_2 = 0$.
We will refer to this scenario as \textit{evaporating capture}.

\section{Dark-matter models that produce annihilation plateaus}
\label{sec:DMmodels_plateau}

The DM models in Sec.~\ref{sec:DMmodels_spike} allow for DM spikes to be formed.
Given the high DM densities in such spikes, they produce the largest DM effects on the binary's evolution, and are likely to be the easiest to distinguish from IMRIs in vacuum. 
However, there are classes of self-annihilating DM models which would decrease the density and flatten the spike's radial density profile; nevertheless, the density in the DM distribution could remain sufficiently high to affect the IMRI's orbit and produce distinctive imprints in the GW data.
For example, a simple model of $s$-wave DM self-annihilation leads to a plateau when the DM-density reaches a ``saturation'' value given by 
\begin{align}
    \rho_\text{sat} &\simeq \frac{m_\DM}{\langle \sigma_\text{an} v\rangle t_\text{BH}} \label{eq:plateau}\\
    &\simeq \frac{\unit[3\times 10^9]{M_\odot}}{\unit{pc^3}} \left(\frac{m_\DM}{\unit{TeV}}\right)\frac{(\unit[3\times 10^{-26}]{cm^3/s})(\unit[10^{10}]{yr})}{\langle \sigma_\text{an} v\rangle t_\text{BH}} , \nonumber
\end{align}
where $t_\text{BH}$ is of order of the age of the BH~\cite{Gondolo:1999ef}.\footnote{Readers might appreciate the relation  $\unit{M_\odot/pc^3} \simeq \unit[38]{GeV/cm^3}$.}
The radius at which the density reaches the saturation value is the annihilation radius, $r_\text{an}$ and, in a naive model, the density is equal to the saturation value at smaller radii.
However, it was shown in~\cite{Vasiliev:2007vh} that having a constant density in this inner region corresponds to having DM particles moving on circular orbits only at each radius within the annihilation radius (rather than a more generic velocity dispersion at this radius).
For isotropic DM spikes with $s$-wave annihilation, Ref.~\cite{Shapiro:2016ypb} proposed truncating the specific energy distribution of the DM particles when it is equal to the gravitational potential of $-G m_1/r_\text{an}$, and showed that inside the annihilation radius, the DM distribution follows a power law $\rho_\DM \propto r^{-\gamma_\text{an}}$ with $\gamma_\text{an} = 1/2$.
For $p$-wave annihilation,~\cite{Shapiro:2016ypb} showed that the density in the annihilation region had a power law of $\gamma_\text{an}\simeq 0.34$.

To encompass the $s$-wave and $p$-wave cases, and to still have a DM distribution with an angular-momentum cutoff at $r_\IN$ for annihilation radii with $r_{\rm in} < r_{\rm an}<r_{\rm sp} $, we use the following analytic form that approximates the numerical results from Ref.~\cite{Shapiro:2016ypb}:
\begin{align}
    \rho_\DM(r)=
\begin{cases}
\rho_{\rm sp}\left(\dfrac{r_{\rm sp}}{r}\right)^{\gamma_{\rm sp}},
& r_{\rm an} \le r \le r_{\rm sp} \,,\\
\rho_{\rm an}\left(\dfrac{r_{\rm an}}{r}\right)^{\gamma_{\rm an}} \left(1-\dfrac{r_\text{in}}{r}\right)^{\gamma_{\rm an}},
& r_{\rm in} \le r < r_{\rm an} \,,\\
0\,,
& r < r_{\rm in}\, .
\end{cases}
\label{eq:rho_ann}
\end{align}
Enforcing continuity at $r_\text{an}$ determines the density $\rho_\text{an}$:
\begin{equation} 
    \rho_\mathrm{an} = \rho_\SP \left(\frac{r_\SP}{r_\mathrm{an}}\right)^{\gamma_\SP} \left(1 - \frac{r_\IN}{r_\mathrm{an}} \right)^{-\gamma_\mathrm{an}} .
\end{equation}
When $r_{\rm in} \ll r_{\rm an} $, then the second term can be neglected, and it can be written as
\begin{equation} \label{eq:rho_an_sat}
    \rho_{\rm an} \simeq \rho_{\rm sp} (r_{\rm sp}/r_{\rm an})^{\gamma_{\rm sp}} \simeq \rho_\text{sat} .
\end{equation}
In this limit, Eq.~\eqref{eq:rho_an_sat} can be solved for the annihilation radius to give
\begin{equation} \label{eq:r_an_rho}
    r_{\rm an} \simeq r_{\rm sp} (\rho_{\rm sp}/\rho_{\rm sat})^{1/\gamma_{\rm sp}} \,.
\end{equation}

\begin{figure}[tb]
    \centering
    \includegraphics[width=0.47\textwidth]{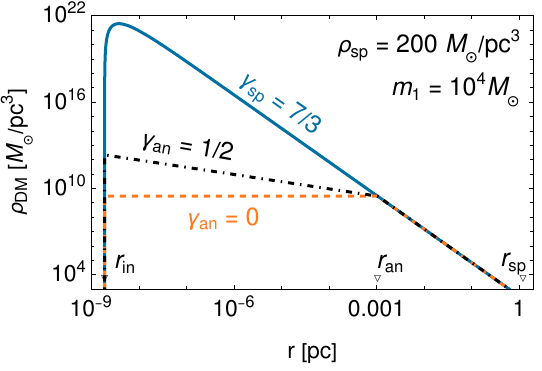}
  \caption{\textbf{DM spikes with and without self-annihilation}.
  All the densities are shown around a primary BH with mass $m_1 = \unit[10^4]{\Msolar}$ (the values of the DM density parameters $\rho_\SP$ and $\gamma_\SP$ are given in the figure).
  The solid blue curve is the Newtonian density with an angular-momentum cutoff used in~\cite{Wade:2025rkk}. 
  The dashed orange curve shows the annihilation plateau from Eq.~\eqref{eq:plateau} with $m_\text{DM}=\unit[1]{TeV}$, $\langle \sigma_\text{an} v\rangle = \unit[3\times 10^{-26}]{cm^3/s}$, and $t_\text{BH} = \unit[10^{10}]{yr}$.
  The dot-dashed black curve is the $s$-wave annihilation case with $\gamma_\text{an} = 1/2$ discussed in~\cite{Shapiro:2016ypb}. 
  For $r> r_\text{sp}$, the profile will transition to the standard DM halo slope (e.g.,~$\gamma\sim 1$ for the Navarro--Frenk--White profile).}
    \label{fig:spike_profile}
\end{figure}

A few example DM densities are illustrated in Fig.~\ref{fig:spike_profile}. 
The height, as well as the onset of the plateau, $r_\text{an}$, is determined in each case by the thermally-averaged DM annihilation cross section $\langle \sigma_\text{an} v\rangle$, which is fixed in some models by the DM abundance.
DM freeze-out models typically have $\langle \sigma_\text{an} v\rangle =\unit[3\times 10^{-26}]{cm^3/s}$ in the early universe~\cite{Cirelli:2024ssz}, but the translation to our situation is model dependent since the DM spike is a different environment from the early universe. Staying agnostic about the DM production history allows us to treat $\langle \sigma_\text{an} v\rangle$ and the plateau density as free model parameters, which could be potentially measurable through GWs.
Depending on the final state, the annihilation cross section can also be subject to strong indirect-detection constraints~\cite{Cirelli:2024ssz}, although again these will be model dependent in general.

In Fig.~\ref{fig:spike_profile}, the values of $\rho_\SP$, $m_1$, and $\gamma_\SP$ are given in the figure, which determine the value of $r_\SP$ from Eq.~\eqref{eq:r_sp}.
The values of $\langle \sigma_\text{an} v\rangle =\unit[3\times 10^{-26}]{cm^3/s}$, $m_\DM=\unit[1]{TeV}$, and $t_\mathrm{BH} = \unit[10^{10}]{yr}$ determine $\rho_\text{sat} \approx \rho_\text{an}$, which determines $r_\mathrm{an}$ from Eq.~\eqref{eq:r_an_rho}.
The solid blue curve shows the Newtonian DM density with an angular-momentum cutoff used in~\cite{Wade:2025rkk}, the black dot-dashed curve is the $s$-wave annihilation profile used in~\cite{Shapiro:2016ypb}, and the dashed orange curve is the less physically-realistic flat plateau.
For the benchmark numbers used in Fig.~\ref{fig:spike_profile}, even cross sections as small as $\langle \sigma_\text{an} v\rangle =\unit[10^{-37}]{cm^3/s}$ would lead to $r_\text{an} > r_\text{in}$ and thus flatten the profile in the innermost region; this  illustrates the potential sensitivity a DM spike could have to even minute annihilation rates.

Having focused on the spike scenario with $\gamma_\text{sp} = 7/3$ for non-annihilating DM in Sec.~\ref{sec:DMmodels_spike}, we will now consider the $s$-wave case with $\gamma_{\rm an} = 1/2$ from Eq.~\eqref{eq:rho_ann} for annihilating DM as a benchmark case to compare against the spike cases in Sec.~\ref{sec:DMmodels_spike}.
We emphasize that for both spikes and plateaus, the respective equations~\eqref{eq:rhoDM} and~\eqref{eq:rho_ann} are the initial DM profiles prior to the binary inspiral; they will be modified by the secondary BH or NS during and after the inspiral.

Similar to the analysis in Sec.~\ref{sec:DMmodels_spike}, we next sketch different classes of DM models that have interactions with NS matter that either does not or does get captured in the NS, and if captured, either remains or evaporates.

\subsection{Self-annihilating dark matter without capture}
\label{sec:plateau}

Dark matter self-annihilation and DM scattering from NS matter need not be related; for example, DM could annihilate into neutrinos or other particles not found in the NS, thereby disentangling the two cross sections.
As one limit of this scenario, we will first assume that the DM-matter cross section is sufficiently small that SA is negligible.
In this scenario, the IMRI will inspiral inside of a DM distribution with a plateau under the influence of gravitational radiation and DF.

However, the value of $r_\mathrm{an}$ in Eq.~\eqref{eq:r_an_rho}, which is determined by the properties of the DM spike and $\rho_\mathrm{sat}$ in Eq.~\eqref{eq:plateau}, can be either inside or outside of the initial orbital radius of the secondary during its inspiral.
As we will describe in more detail in Sec.~\ref{subsubsec:diffs}, DF begins weakening when the binary passes within the annihilation radius, because there are fewer DM particles moving more slowly than the speed of the secondary, which are those that contribute to the DF force.
This, combined with the fact that DM self-annihilation decreases the density overall, implies that the annihilation radius must be sufficiently small (or $\rho_\mathrm{sat}$ sufficiently large) to be able to distinguish an IMRI in an DM annihilation plateau from a vacuum system.
We will show results in Sec.~\ref{sec:results} for different radii $r_\mathrm{an}$ that illustrate where DM annihilation plateaus and DM spikes are likely to have distinguishable GW dephasings; we consider a similar comparison for IMRIs in DM plateaus and in vacuum.

\subsection{Self-annihilating dark matter with capture}
\label{sec:plateau+SA}

Next, we discuss the typical case of weakly-interacting DM, in which DM can annihilate \emph{and} has large-enough scattering cross sections to enable capture by the NS. 
The captured DM might thermalize with the NS matter and eventually start annihilating inside the NS, heating it up~\cite{Bramante:2023djs}. 
For simplicity, we will assume that the thermalization timescale is sufficiently long compared to the inspiral so that annihilation inside the NS (or at least the eventual radiation out of the NS) is inefficient.
In this case, DM capture still leads to a NS mass growth according to Eq.~\eqref{eq:accretion}. 

There will be no significant change in dynamical friction from the scenarios discussed in Sec.~\ref{sec:plateau}.
Whether this class of DM models can be distinguished from that in Sec.~\ref{sec:plateau} will depend on the size of the effects of SA and the corresponding mass increase on the binary dynamics.
This will be discussed in Sec.~\ref{sec:results}.
In terms of distinguishing a BH from a NS secondary, DF and SA for the NS will be quantitatively similar to a secondary BH with the same mass; the main difference in this case is that the merger takes place in a DM plateau rather than a DM spike.
Because the plateau decreases the density, DM effects on the IMRI will be suppressed.
We expect that it will be more difficult to distinguish BH secondaries from NS secondaries in this scenario.

\subsection{Annihilating DM with evaporating capture}
\label{sec:plateau+SA+no_growth}

Building on the previous case, we now consider efficient DM annihilation inside the NS into particles that escape the NS. 
As noted in Sec.~\ref{sec:spike+SA+no_growth}, examples of such particles are low-energy neutrinos or other weakly-interacting particles. 
Since the accumulated DM component inside the NS is now evaporating away, the NS mass increase is suppressed compared to the usual capture scenarios. 
In the same extreme limit considered in Sec.~\ref{sec:spike+SA+no_growth}, the NS would not accumulate mass at all: $\dot{m}_2 = 0$.
It will still experience a drag force from the one-sided DM collisions, which produces a change in the orbital separation given by $\dot{r}_2 \simeq -2 r_2 \rho_\DM \sigma_\text{NS} v_2/m_2$.

The largest difference between the scenarios of evaporating capture in a DM plateau and the capture without evaporation in  Sec.~\ref{sec:plateau+SA} is the fact that the chirp mass is constant during the inspiral rather than evolving.
Given the high precision with which the chirp mass can be measured (see, e.g.,~\cite{Coogan:2021uqv}) even relative changes in the chirp mass of order $10^{-4}$ could be distinguished from a constant chirp mass, so there is the possibility that the evaporating capture scenario could be distinguished from the ordinary capture case in Sec.~\ref{sec:plateau+SA}.
Because BH secondaries do not have evaporating capture, this also makes this scenario more amenable to distinguishing a BH from a NS as the small compact object (compared with that in~\ref{sec:plateau+SA}).
However, because the increase in mass speeds up the inspiral, it produces larger dephasing effects from vacuum systems where the secondary mass is fixed; in this sense it could be less favorable to distinguish the evaporating capture case from that of vacuum IMRIs.

\section{Joint evolution of the DM density and binary} \label{sec:evolution}

In the discussion below, we will use $m_2$ for the mass of the secondary compact object in the IMRI (recall $m_1$ is the mass of the primary black hole).
We will make the approximations used in~\cite{Nichols:2023ufs,Wade:2025rkk} that the mass ratio is $q=m_2/m_1 \ll 1$, and we will work to leading order in $q$.
For example, the total mass and reduced mass will satisfy
\begin{equation}
    M \equiv m_1 + m_2 \approx m_1, \quad \mu \equiv \frac{m_1 m_2}M \approx m_2
\end{equation}
We will use $v$ and $r$ for the speed and position of a dark matter particle and $v_2$ and $r_2$ for those quantities for the secondary.

\subsection{Dark-matter distribution and density} \label{subsec:DMdensity}

In this paper, we will consider dark matter distributions that are spherically symmetric and isotropic, so that they can be described in terms of the specific relative energy of a DM particle in the gravitational potential of the massive BH:
\begin{equation}
    \mathcal E = \frac{Gm_1}{r} - \frac 12 v^2 = \Phi(r) - \frac 12 v^2
\end{equation}
The distribution function will be denoted by $f_{(i)}(\mathcal E,t)$, where the particular value of the subscript $(i)$ denotes whether it is associated with a spiky or a plateau DM distribution (both of which will have an angular-momentum cutoff, which is discussed below).
The DM density in position space is obtained by integrating $f(\mathcal E,t)$ over the permitted velocities at a given radius $r$: namely,
\begin{equation} \label{eq:rho_from_f}
    \rho_\DM^{(i)}(r,t) = \int_{v_\MIN}^{v_\MAX} \ud^3v \, f_{(i)}(\mathcal E,t) .
\end{equation}
Here $v_\MIN$ and $v_\MAX$ are the minimum and maximum velocities, which will depend on what case $i$ for the DM density is being considered.
These velocities are also functions of radius $r$.
Alternately, the integral in Eq.~\eqref{eq:rho_from_f} can be rewritten in terms of the energy at fixed $r$, by using the facts that $\ud^3 v = 4\pi v^2 \ud v$ for isotropic distributions and $\ud\mathcal E = -v \ud v$:
\begin{equation} \label{eq:rho_from_f_E}
    \rho_\DM^{(i)}(r,t) = 4\pi \int_{\mathcal E_\MIN}^{\mathcal E_\MAX} \ud \mathcal E \sqrt{2\left(\frac{Gm_1}{r} - \mathcal E \right)} f_{(i)}(\mathcal E,t) .
\end{equation}
At a fixed $r$, $\mathcal E_\MIN$ is $\mathcal E$ evaluated at $v_\MAX$, and vice versa for $\mathcal E_\MAX$.
To compute the full density in Eq.~\eqref{eq:rho_from_f_E}, we will choose the minimum energy to be $\mathcal E_\MIN = 0$, which neglects DM particles on unbound orbits.

In~\cite{Kavanagh:2020cfn,Nichols:2023ufs}, the maximum energy considered was given by
\begin{equation}
    \mathcal E_\MAX = \mathcal E_\MAX^{(\Phi)}(r) \equiv \frac{Gm_1}r
\end{equation}
However, in~\cite{Wade:2025rkk}, a minimum angular momentum of $\Jmin = Gm_1 j_\MIN/c$, where $j_\MIN = \sqrt{8}$ was introduced.
This limited the maximum energy to be smaller:
\begin{equation} \label{eq:EmaxJ}
    \mathcal E_\MAX = \mathcal E_\MAX^{(\mathcal J)}(r) \equiv \frac{Gm_1}r - \frac{\Jmin^2}{2r^2} .
\end{equation}
For this choice, the density smoothly went to zero at $r = r_\mathcal{J}/2$, where $r_\mathcal{J} = \Jmin^2/(Gm_1)$, and it was consistent with the fact that DM particles with angular momentum lower than this value would be captured by the primary $m_1$.
The case $i = \mathcal J$ with the angular-momentum cutoff will be what we refer to as a DM spike in this paper.
The spike with a position-space cutoff ($i = \Phi$) will not be used in the simulations in this paper; we introduced it primarily for describing the relationship of our approach to that used in other works.

\subsubsection{Differences for self-annihilating DM} \label{subsubsec:diffs}

The works~\cite{Kavanagh:2020cfn,Nichols:2023ufs,Wade:2025rkk} all assumed that there was no DM-DM annihilation.
However, as discussed in~\cite{Shapiro:2016ypb} (see also~\cite{Vasiliev:2007vh}), the effects of DM-DM annihilation in the case $j_\MIN=0$ can be implemented by imposing a maximum energy of $\mathcal E_\MAX = \mathcal E_\MAX^{(\Phi)}(r_\mathrm{an})$ on the DM particles for $r < r_\mathrm{an}$ and $\mathcal E_\MAX = \mathcal E_\MAX^{(\Phi)}(r)$ for $r \geq r_\mathrm{an}$.
Here $r_\mathrm{an}$ is the ``annihilation radius.''
The rationale behind this cutoff was that a DM particle with energy greater than this value would need to be on an orbit completely within $r_\mathrm{an}$, which would make it likely that the DM particle would annihilate.
With a nonzero $j_\MIN$, the combined effects of DM-DM annihilation and capture of low angular-momentum DM particles into the primary can be accounted for through an appropriate set of conditions involving $\mathcal E_\MAX^{(\mathcal J)}(r)$ (see also~\cite{Shapiro:2016ypb}).

First, it is useful to note that $\mathcal E_\MAX^{(\mathcal J)}(r)$ has a peak at $r=r_\mathcal{J}$, and that the equation $\mathcal E_\MAX^{(\mathcal J)}(r) = \mathcal E_\MAX^{(\mathcal J)}(r_\mathrm{an})$ has two solutions: the ``obvious'' one at $r=r_\mathrm{an}$ and one at what will be denoted $r = r'_\mathrm{an}$, which is given by
\begin{equation} \label{eq:r_an_prime}
    r'_\mathrm{an} = \frac{r_\mathcal{J} r_\mathrm{an}}{2r_\mathrm{an} - r_\mathcal{J}} .
\end{equation}
It follows that when there is DM-DM annihilation and an angular-momentum cutoff, the maximum energy used at the upper integration limit in Eq.~\eqref{eq:rho_from_f_E} should be 
\begin{align} \label{eq:EmaxJan}
    \mathcal E_\MAX & = \mathcal E_\MAX^{(\mathcal J+\mathrm{an})}(r) \nonumber \\
    & \equiv
    \begin{cases}
        \mathcal E_\MAX^{(\mathcal J)}(r_\mathrm{an}) & \text{if } \, r_\mathrm{an} > r_\mathcal{J} , \quad r'_\mathrm{an} < r < r_\mathrm{an} ,\\
        \mathcal E_\MAX^{(\mathcal J)}(r) & \text{otherwise} ,
    \end{cases}
\end{align}
for $ \mathcal E_\MAX^{(\mathcal J)}(r)$ given in Eq.~\eqref{eq:EmaxJ}.

For computing dynamical friction, we will be interested in the density of particles at $r_2$ that are moving more slowly than the orbital speed of the secondary,  $v_2 = \sqrt{Gm_1/r_2}$.
As in~\cite{Wade:2025rkk}, we will denote this density by $\rho_\DM^{(i)}(r_2,t;v<v_2)$, and it can be obtained from the same integral in Eq.~\eqref{eq:rho_from_f_E} with $r=r_2$ and with $\mathcal E_\MIN = Gm_1/(2r_2)$.

With an angular momentum cutoff, it was noted in~\cite{Wade:2025rkk} that this $\mathcal E_\MIN$ is less than the upper limit of the integral when $Gm_1/(2r_2) < \mathcal E_\MAX^{(\mathcal J)} (r_2)$ which occurs for $r_2 > r_\mathcal{J}$.
An implication of this is that dynamical friction weakens as the secondary approaches $ r_\mathcal{J}$ and ``turns off'' for $r_2 <  r_\mathcal{J}$, because the density that contributes to dynamical friction, $\rho_\DM^{(i)}(r_2,t;v<v_2)$, goes to zero for $r_2 <  r_\mathcal{J}$. 
With DM-DM annihilation and with $r_\mathrm{an} > r_\mathcal{J}$, Eq.~\eqref{eq:EmaxJan} shows the maximum energy $\mathcal E_\MAX^{(\mathcal J)}(r_\mathrm{an})$, is smaller than $\mathcal E_\MAX^{(\mathcal J)}(r_2)$ for $r_2 \in (r'_\mathrm{an},r_\mathrm{an})$.
This further decreases the density $\rho_\DM^{(i)}(r_2,t;v<v_2)$ relevant for dynamical friction.

By computing the values of $r_2$ for which $Gm_1/(2r_2) < \mathcal E_\MAX^{(\mathcal J)} (r_\mathrm{an})$ holds, it follows that the density $\rho_\DM^{(i)}(r_2,t;v<v_2)$ is nonzero for 
\begin{equation} \label{eq:r2_DF_cut}
    r_2 > \frac{r_\mathrm{an} r'_\mathrm{an}}{r_\mathcal{J}} = \frac{r_\mathrm{an}^2}{2r_\mathrm{an} - r_\mathcal{J}} .
\end{equation}
The equality was obtained using the expression for $r'_\mathrm{an}$ in Eq.~\eqref{eq:r_an_prime}.
The quantity $r_\mathrm{an} r'_\mathrm{an}/r_\mathcal{J}$ is greater than $r_\mathcal{J}$ and equal only when $r_\mathrm{an} = r_\mathcal{J}$.
Thus, having DM-DM annihilation will cause dynamical friction to turn off at a larger separation than it does with just the angular momentum cutoff.

Note that the density $\rho_\DM^{(i)}(r_2,t;v<v_2)$ will begin decreasing from the case without DM-DM annihilation at $r_2 = r_\mathrm{an}$, because of the form of the maximum energy in Eq.~\eqref{eq:EmaxJan}.
Thus, there will be a range of radii $r_2 \in (r_\mathrm{an} r'_\mathrm{an}/r_\mathcal{J}, r_\mathrm{an})$ where dynamical-friction force on the secondary in a spike with DM-DM annihilation will be weaker (though nonzero) than the case with no annihilation.
Because $r_\mathrm{an} r'_\mathrm{an}/r_\mathcal{J}$ is greater than $r_\mathcal{J}$ for $r_\mathrm{an} > r_\mathcal{J}$, the DF force will be zero for $r_2 \leq r_\mathrm{an} r'_\mathrm{an}/r_\mathcal{J}$.

\subsection{Binary and DM evolution equations}

We first discuss the evolution equations for the secondary.
We focus on binaries in circular orbits, as in~\cite{Wade:2025rkk}, and we consider the evolution of the binary's separation $r_2(t)$ on timescales longer than the orbital timescale.
Next, we discuss the evolution of the DM distribution function in response to feedback from the binary's interaction with the DM.
Much of the discussion will be a review of the formalism in~\cite{Wade:2025rkk}, and we refer the reader to~\cite{Wade:2025rkk} for further details.
We devote more text to the differences between this work and that of~\cite{Wade:2025rkk}.

\subsubsection{Binary evolution equations}

Similar to~\cite{Wade:2025rkk}, we write the evolution equation for $r_2(t)$ as
\begin{subequations} \label{eq:r2evolution}
    \begin{equation} \label{eq:dot_r2_gen}
        \dot r_2 = -\dot r_2^\mathrm{RR} - \dot r_{2,(i)}^\mathrm{DF} - \delta_{(1)}^{(j)} \dot r_{2,(i)}^\mathrm{SA}
    \end{equation}
    where unlike in~\cite{Wade:2025rkk} we introduced a Kronecker delta $\delta_{(1)}^{(j)}$, which eliminates the effects of secondary accretion in the case $j=0$ and incorporates it when $j=1$.
    We also labeled the dynamical friction and secondary accretion forces with the index $i$ that distinguishes the different cases for the DM density in Sec.~\ref{subsec:DMdensity} (specifically, ``$\mathcal J$'' or ``$\mathcal J+\mathrm{an}$'').
    As in~\cite{Wade:2025rkk}, we work to leading order in the mass ratio $q=m_2/m_1$ where the different terms on the right-hand side of Eq.~\eqref{eq:dot_r2_gen} are given by
    \begin{align}
        \dot r_2^\mathrm{RR} & = \frac{64} 5 q c \left(\frac{Gm_1}{c^2 r_2} \right)^3 , \\
        \dot r_{2,(i)}^\mathrm{DF} & = 8\pi q \sqrt{\frac{G}{m_1}} (\log \Lambda) r_2^{5/2} \rho_\DM^{(i)}(r_2,t; v< v_2) , \\
        \dot r_{2,(i)}^\mathrm{SA} & = \frac 2{m_2 } \sqrt{Gm_1 r_2} \sigma_l(v_2) \rho_\DM^{(i)}(r_2,t) .
    \end{align}
\end{subequations}
The label $l$ can be BH for the cross section in Eq.~\eqref{eq:BH_cross_section} or NS for the cross section in Eq.~\eqref{eq:sigma_NS}.
When accretion is present, the mass $m_2$ will increase.
However, as discussed in Sec.~\ref{sec:DMmodels_spike}, it is also possible for DM to annihilate in the NS (evaporating capture).
This leaves the drag force from secondary accretion in Eq.~\eqref{eq:dot_r2_gen}, but changes the evolution of the mass to be of the form
\begin{equation} \label{eq:dot_m2_gen}
    \dot m_2^{(k)} = \delta_{(0)}^{(k)} \sigma_l(v_2) \rho_\DM^{(i)}(r_2,t) v_2 .
\end{equation}
As in Eq.~\eqref{eq:dot_r2_gen}, we introduce a Kronecker delta, now $\delta_{(0)}^{(k)}$, which corresponds to no annihilation in the secondary ($k=0$) or complete annihilation ($k=1$).

To specify a physical scenario described in Secs.~\ref{sec:DMmodels_spike} and~\ref{sec:DMmodels_plateau}, we will choose a secondary type ($l=$BH or NS), and a triple of cases $(i,j,k)$.
For BH secondaries, we will consider just a reference case given in~\cite{Wade:2025rkk}: namely, $(i,j,k)=(\mathcal J, 1, 0)$.
This configuration has the angular-momentum cutoff in the DM density, secondary accretion present, and no annihilation of DM particles in the secondary.
For NS secondaries, we consider both the $i=\mathcal J$ and $\mathcal J+\mathrm{an}$ cases, secondary accretion and not ($j=1$ and 0), and in the case that $j=1$, both no annihilation and annihilation in the secondary ($k=0$ and 1).
This encompasses the six scenarios discussed in Secs.~\ref{sec:DMmodels_spike} and~\ref{sec:DMmodels_plateau}.

\subsubsection{Evolution of the DM distribution function}

We now review some aspects of the evolution of the distribution function.
First, we will again make use of the density of states,
\begin{equation}
    g_{(i)}(\mathcal E) = \int \ud^3r \int \ud^3 v \, \delta \boldsymbol(\mathcal E - \mathcal E(r,v) \boldsymbol ) . 
\end{equation}
As in~\cite{Wade:2025rkk}, in the case of an angular momentum cut without DM-DM annihilation, the integral can be evaluated to give
\begin{equation}
    g_{(\mathcal J)}(\mathcal E) = \sqrt 2 (\pi G m_1)^3 \mathcal E^{-5/2} \left(1 - \frac{2\mathcal E}{v_\mathcal{J}^2} \right) 
\end{equation}
for energies in the interval $\mathcal E \in (0, Gm_1/(2r_\mathcal{J})]$ (otherwise it is zero).
Here $v_\mathcal{J}^2 = Gm_1/r_\mathcal{J}$ was defined as the velocity squared of a circular orbit at $r=r_\mathcal{J}$.
With DM-DM annihilation, the expression is modified to
\begin{equation}
    g_{(\mathcal J+\mathrm{an})}(\mathcal E) =
    \begin{cases}
         g_{(\mathcal J)}(\mathcal E) & \text{if } \, 0 < \mathcal E \leq \mathcal E_\MAX^{(\mathcal J)}(r_\mathrm{an}) , \\
         0 & \text{otherwise}.
    \end{cases}
\end{equation}

The density of states $ g_{(i)}(\mathcal E)$ enters into the rate coefficients that determine the evolution of the distribution function $f_{(i)}(\mathcal E, t)$.
The rate coefficient for accretion has the same form as that in~\cite{Wade:2025rkk} when the maximum energy is replaced with the appropriate maximum energy for the relevant case
\begin{align} \label{eq:RcalEacc}
    & R_{\mathcal E}^{(i),\mathrm{acc}} = \nonumber \\
    & \begin{cases}
        \dfrac{8\pi^2 r_2 \sigma_l(v_2)}{T_2 g_{(i)}(\mathcal E)} \sqrt{2[\mathcal E_\MAX^{(i)}(r_2)-\mathcal E]} & \text{if } \, 0 < \mathcal E \leq \mathcal E_\MAX^{(i)}(r_2) , \\
        0 & \text{otherwise}.
    \end{cases}
\end{align}
Here we introduced $T_2 = 2\pi \sqrt{r_2^3/(Gm_1)}$ as the orbital period of the secondary.
In the case without DM-DM annihilation, the cross section $\sigma_l(v_2)$ needs to be modified from those in Eqs.~\eqref{eq:BH_cross_section} and~\eqref{eq:sigma_NS} in the limit that $\mathcal E$ approaches the largest DM particle energy in the distribution function, $Gm_1/(2r_\mathcal{J})$.
The change is necessary, because in Eq.~\eqref{eq:RcalEacc}, the density of states goes to zero as $\mathcal E \rightarrow Gm_1/(2r_\mathcal{J})$, but the cross section remains finite, thereby causing a divergence in $R_\mathcal{E}^{\mathcal{(J)},\mathrm{acc}}$ at this largest possible energy.
This divergence was not physical, but rather related to the approximation used in computing $R_\mathcal{E}^{\mathcal{(J)},\mathrm{acc}}$, which breaks down when the energy approaches the upper limit $\mathcal E_\MAX$.

The resolution in~\cite{Wade:2025rkk} was to introduce an accretion impact parameter, which for the BH and NS cases is given by
\begin{equation}
    \left \{ 
    \begin{array}{ll}
       b^\mathrm{acc}_\mathrm{BH} & = \dfrac{4Gm_2}{c v_2} , \\
       b^\mathrm{acc}_\mathrm{NS} & =\dfrac{v_\mathrm{esc} R_\mathrm{NS} }{v_2\sqrt{1 - v_\mathrm{esc}^2/c^2}} ,
    \end{array}
    \right.
\end{equation}
respectively.
Next, an energy-dependent cross section was defined that reduces to the energy-independent value for most energies, but vanishes as $\mathcal E \rightarrow Gm_1/(2r_\mathcal{J})$.
Its definition requires introducing the radii $r_\pm$ in~\cite{Wade:2025rkk}, which are the radii at which the density of states vanishes for a given energy:
\begin{equation}
    r_\pm = \frac{Gm_1}{2\mathcal E} \left(1 \pm \sqrt{1 - \frac{2\mathcal E}{v_\mathcal{J}^2}} \right) .
\end{equation}
From these radii and the accretion impact parameter an angle was defined by
\begin{equation}
    \theta_l = \cos^{-1}[\min\boldsymbol(1, (r_+-r_-)/(2b^\mathrm{acc}_l) \boldsymbol)] ,
\end{equation}
and the corresponding accretion cross section was given by
\begin{equation} \label{eq:sigma_finite}
    \sigma_l(v_2) = [\pi - 2\theta + \sin(2\theta)] (b_l^\mathrm{acc})^2 .
\end{equation}
Note, however, that for $r_\mathrm{an} > r_{\mathcal J}$, the density of states has a nonzero value as $\mathcal E \rightarrow \mathcal E_\MAX^{(\mathcal J )}(r_\mathrm{an})$, so the accretion rate $R_\mathcal{E}^{\mathcal{(J}+\mathrm{an}),\mathrm{acc}}$ remains finite without requiring the prescription used in Eq.~\eqref{eq:sigma_finite}.
Nevertheless, we use it in both the $\mathcal J$ and $\mathcal J + \mathrm{an}$ cases.

The final element needed to specify the evolution of the DM distribution is the differential scattering rate $\mathcal R^{(i)}_\mathcal{E}(\Delta \mathcal E)$ per scattering energy transfer $\Delta \mathcal E$.
The total scattering rate is given by the integral over all permitted scattering energy changes $\Delta \mathcal E$:
\begin{equation}
    R^{(i)}_{\mathcal E} = \int \ud (\Delta \mathcal E) \mathcal R^{(i)}_{\mathcal E} (\Delta \mathcal E) .
\end{equation}
We give the expression for $\mathcal R^{(\mathcal J+\mathrm{an})}_\mathcal{E}(\Delta \mathcal E)$, which is given in terms of $\mathcal R^{(\mathcal J)}_\mathcal{E}(\Delta \mathcal E)$, which is the expression computed in~\cite{Wade:2025rkk}:
\begin{equation}
    \mathcal R^{(\mathcal J+\mathrm{an})}_\mathcal{E}(\Delta \mathcal E) = 
    \begin{cases}
        \mathcal R^{(\mathcal J)}_\mathcal{E}(\Delta \mathcal E) & \text{if } \, 0 < \mathcal E \leq \mathcal E_\MAX^{(\mathcal J)}(r_\mathrm{an}) , \\
        0 & \text{otherwise} .
    \end{cases}
\end{equation}
The expression for $R^{(\mathcal J)}_\mathcal{E}(\Delta \mathcal E)$ is lengthy, but it arises from evaluating the integral
\begin{align} \label{eq:calRdeltaE}
    \mathcal R^{(\mathcal J)}_\mathcal{E}(\Delta \mathcal E) = \frac{4\pi}{T_2 g(\mathcal E)}  \int \ud^3 r \int \ud v & \sqrt{v^2 - v_\MIN^2} \delta \boldsymbol( \mathcal E - \mathcal E(r,v) \boldsymbol) \nonumber \\
    & \times \delta \boldsymbol(\Delta \mathcal E(b) - \Delta \mathcal E \boldsymbol) .
\end{align}
The change in energy in a scattering event as a function of impact parameter was approximated in~\cite{Wade:2025rkk} by
\begin{equation}
    \Delta \mathcal E(b) = -2 v_2^2 \left[1 + \left(\frac{b}{b_{90}}\right)^2 \right]^{-1} ,
\end{equation}
where $b_{90}$ is the impact parameter for a $90^\circ$ scattering angle, and the approximation made was using $v_2$ for the relative speed of the encounter between the secondary and the DM particle.
The full expression for the result of the integral in Eq.~\eqref{eq:calRdeltaE} can be obtained from Eqs.~(A3)--(A12) of~\cite{Wade:2025rkk}.

The evolution of the dark matter distribution function $f_{(i)}(\mathcal E, t)$ can now be written in terms of the accretion rate $R^{(i),\mathrm{acc}}_{\mathcal E}$ (per orbit of the secondary), the total dynamical-friction scattering rate $R^{(i)}_{\mathcal E}$, the differential rate $\mathcal R^{(i)}_{\mathcal E}$, and the ratio of the densities of states at two energies,
\begin{equation}
    h_{(i)}(\mathcal E, \Delta \mathcal E) \equiv \frac{g_{(i)}(\mathcal E - \Delta \mathcal E)}{g_{(i)}(\mathcal E)} .
\end{equation}
The integral-differential equation is
\begin{align} \label{eq:f_evolve}
    \frac{\partial f_{(i)}}{\partial t} = & - (R^{(i)}_{\mathcal E} + R^{(i),\mathrm{acc}}_{\mathcal E}) f_{(i)}(\mathcal E, t) \nonumber \\
    & + \int \ud(\Delta \mathcal E) h_{(i)}(\mathcal E, \Delta \mathcal E) \mathcal R^{(i)}_{\mathcal E - \Delta\mathcal E}(\Delta \mathcal E) f_{(i)}(\mathcal E - \Delta \mathcal E, t) .
\end{align}
This equation is coupled to the dynamics of the binary in Eq.~\eqref{eq:r2evolution}, because the rate coefficients are functions of $r_2$, which is most easily seen in the expression for $R^{(i),\mathrm{acc}}_{\mathcal E}$ in Eq.~\eqref{eq:RcalEacc}, but is implicit in the expression for $ \mathcal R^{(i)}_\mathcal{E}(\Delta \mathcal E)$ in Eq.~\eqref{eq:calRdeltaE}.
The evolution equations for $r_2$ depend explicitly on $f_{(i)}(\mathcal E, t)$ via the fact that the density is given by the integral of the distribution function in Eq.~\eqref{eq:rho_from_f_E}.
The evolution of $f_{(i)}(\mathcal E,t)$ and $r_2$ must be solved as a coupled set of integral-differential equations.

\section{Simulation methods, initial conditions, and observables}
\label{sec:methods}

In this section, we discuss our simulations of the coupled IMRI–DM equations of motion with either a BH or NS secondary.
We will review some aspects of the implementation of the IMRI-DM evolution equations in the \textsc{HaloFeedbackAcc} code~\cite{HaloFeedbackAcc}, the choice of initial conditions, and the metrics used for analyzing the simulation outputs.
The results for the GW dephasing and DM density will be presented in Sec.~\ref{sec:results}.

\subsection{Simulation methods and initial data}

We use a modified version of the \textsc{HaloFeedbackAcc} code~\cite{HaloFeedbackAcc} (which was based on the original \textsc{HaloFeedback} code~\cite{HaloFeedback}) to perform the simulations described in this paper.
Many of the details of running the simulations are similar to those described in~\cite{Nichols:2023ufs, Wade:2025rkk}, although some modifications were required to implement the changes discussed in Sec.~\ref{sec:evolution}.
For example, we implemented into~\cite{HaloFeedbackAcc} the different cases of DM densities in Eq.~\eqref{eq:rho_from_f}, the cross section appropriate for both BH and NS secondaries in Eq.~\eqref{eq:sigma_finite}, the different DM scenarios in Secs.~\ref{sec:DMmodels_spike} and~\ref{sec:DMmodels_plateau} that add or remove the terms in the evolution equations for $\dot r_2$ and $\dot m_2$ in Eqs.~\eqref{eq:dot_r2_gen}, and the modified SA and DF (differential) scattering rates in Eqs.~\eqref{eq:RcalEacc} and~\eqref{eq:calRdeltaE}.\footnote{As a matter of practical code implementation, the changes to the computation of the density and the effect of the changes to the scattering rates in the evolution equation for the DM distribution function $f_{(i)}(\mathcal E,t)$ can be implemented most straightforwardly by setting $f_{(i)}(\mathcal E,t)=0$ for energies $\mathcal E$ outside of the interval $[\mathcal E_\MIN,\mathcal E_\MAX]$, as in Eq.~\eqref{eq:f_J_an}.}

We simulate the inspiral for two (initial) mass ratios: $q = 10^{-4}$ and $q = 10^{-3}$.
Specifically, we fix the initial mass of the secondary to be $m_{2,0} = 1.4~\Msolar$ (typical of a canonical galactic NS) and set the primary mass $m_1$ to be either $m_1 = \unit[1.4 \times 10^4]{\Msolar}$ or $\unit[1.4 \times 10^3]{\Msolar}$, respectively.
For the initial spike DM density profile in Eq.~\eqref{eq:rhoDM}, we use $\rho_\SP = \unit[200]{\Msolar/pc^3}$ and $\gamma_\SP = 7/3$ (and $j_\mathrm{min} = \sqrt 8$ to determine $r_\IN$).
The initial distribution function $f_{(i)}(\mathcal E,0)$ that reproduces this initial density was shown in~\cite{Wade:2025rkk} to be given by
\begin{equation}
    f_{(\mathcal{J})}(\mathcal E,0) = \frac{\gamma_\SP(\gamma_\SP-1)\Gamma(\gamma_\SP-1)}{(2\pi)^{3/2}\Gamma(\gamma_\SP-1/2)} \left(\frac{r_\SP \mathcal E}{G m_1}\right)^{\gamma_\SP} \! \rho_\SP \mathcal E^{-3/2} .
\end{equation}
For the plateau DM density in Eq.~\eqref{eq:rho_ann}, we use an initial distribution function given by
\begin{equation} \label{eq:f_J_an}
    f_{(\mathcal J+\mathrm{an})}(\mathcal E,0) = 
    \begin{cases}
        f_{(\mathcal J)}(\mathcal E,0) & \text{if } \, 0 < \mathcal E \leq \mathcal E_\MAX^{(\mathcal J)}(r_\mathrm{an}) , \\
        0 & \text{otherwise} ,
    \end{cases}
\end{equation}
We use the same values of $\rho_\SP$ and $\gamma_\SP$,  which enter into $f_{(\mathcal J)}(\mathcal E,0)$, as in the DM spike case.

\begin{figure}
    \centering
    \includegraphics[width=\columnwidth]{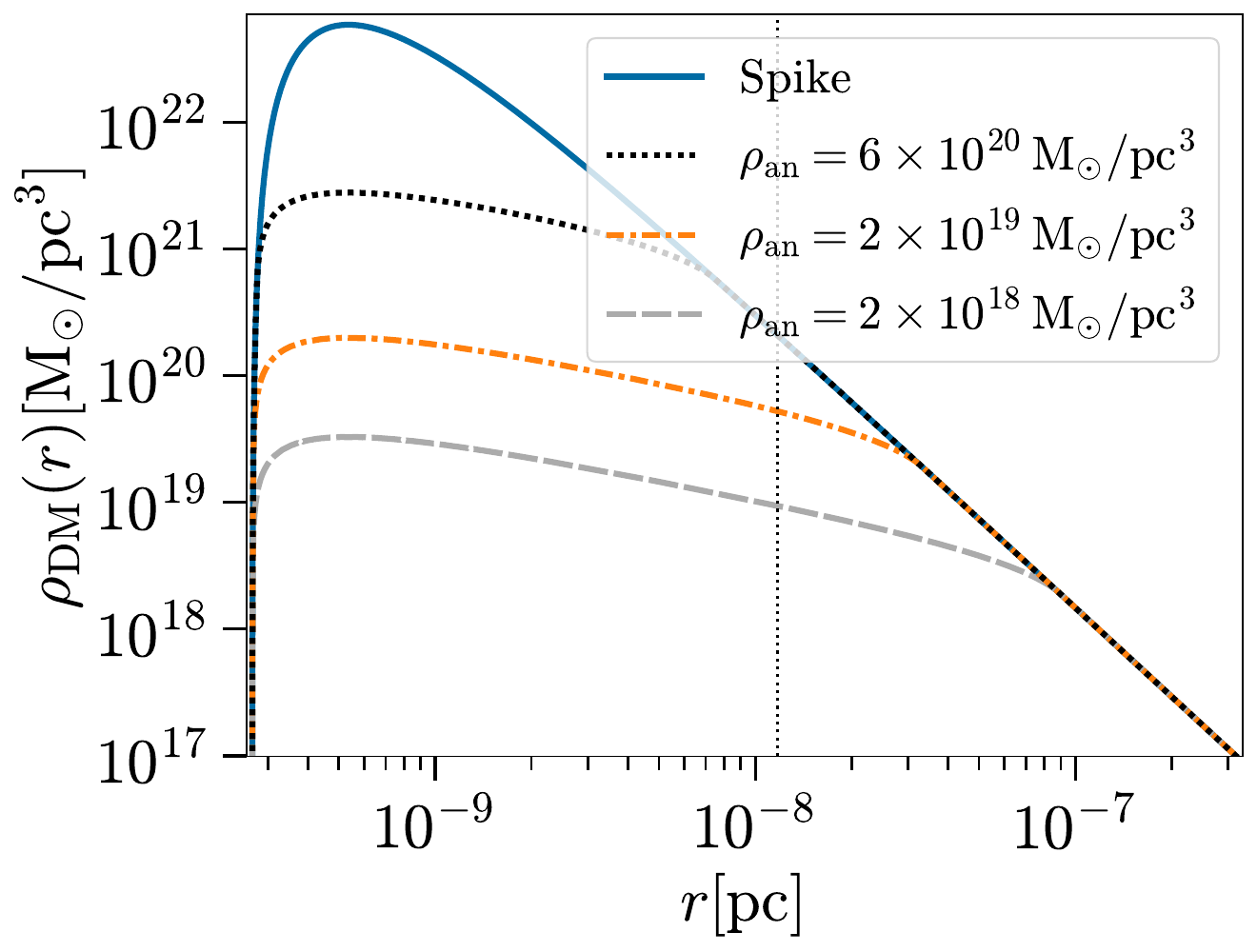} \
    \includegraphics[width=\columnwidth]{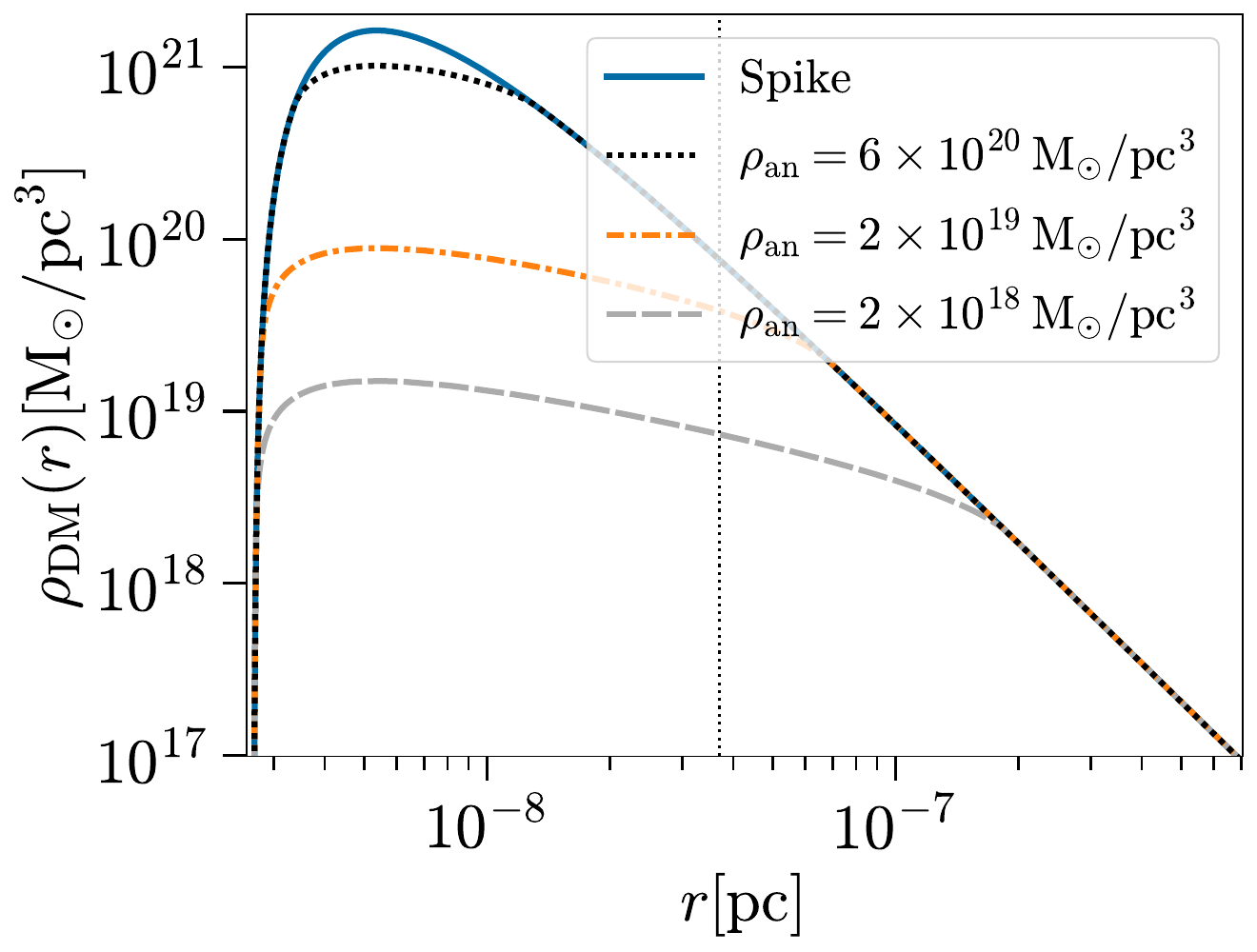}
    \caption{\textbf{Initial DM density with and without self-annihilation around two different primary BHs}. 
    Both panels show the initial DM densities used in the simulations.
    The spiked profile is shown as a solid blue curve, and the three annihilation plateau profiles with different saturation densities are $\rho_\mathrm{an} = 2\times 10^{18} \, \Msolar/\mathrm{pc}^3$ in dashed gray, $\rho_\mathrm{an} = 2\times 10^{19} \, \Msolar/\mathrm{pc}^3$ in dash-dotted orange, $\rho_\mathrm{an} = 6\times 10^{20} \, \Msolar/\mathrm{pc}^3$ in dotted black.
    As noted in Sec.~\ref{sec:methods}, we choose $\rho_\SP = 200\,\Msolar \, \mathrm{pc}^{-3}$ and $\gamma_\SP = 7/3$ in all cases.
    The primary BH mass is $m_1 = 1.4 \times 10^3\,\Msolar$ (\emph{top}) and $m_1 = 1.4 \times 10^4\,\Msolar$ (\emph{bottom}).
    A thin, vertical, dotted line is placed at the value of $r_\fy$ computed with $m_1$ and $m_{2,0}$.
    }
    \label{fig:density_initial}
\end{figure}

To fully specify the distribution function $f_{(\mathcal J+\mathrm{an})}(\mathcal E,0)$, we also need to select values of $r_\mathrm{an}$.
To do so, we select values of $\rho_\mathrm{an}$ and use Eqs.~\eqref{eq:rho_an_sat} and \eqref{eq:r_an_rho} to determine the corresponding values of $r_\mathrm{an}$.
The largest values of $r_\mathrm{an}$ (lowest $\rho_\mathrm{an}$) were selected empirically to produce an inspiral around the threshold of what could be distinguished from an inspiral in vacuum (for the more massive $m_1$ in case~\ref{sec:plateau+SA+no_growth}).
This corresponds to a $\rho_\mathrm{an} \approx \rho_\mathrm{sat}$ of $\unit[2\times 10^{18}]{\Msolar/pc^3}$.
We also choose two other values of $\rho_\mathrm{an}$ such that the annihilation radius $r_\mathrm{an}$ lies outside or inside the radius $r_\fy$, which is the radius for which the IMRI reaches the innermost stable circular orbit (ISCO) radius in four years.
The values of these densities are $\rho_\mathrm{an} = \unit[2\times 10^{19}]{\Msolar/pc^3}$ and $\rho_\mathrm{an} = \unit[6\times 10^{20}]{\Msolar/pc^3}$.
The densities for a DM spike and a DM plateau with these three values of $\rho_\mathrm{an}$ are shown in Fig.~\ref{fig:density_initial}.
The top panel is the density around a BH of mass $m_1 = 1.4 \times 10^3\,\Msolar$ and the bottom panel is for $m_1 = 1.4 \times 10^4\,\Msolar$.

The \textsc{HaloFeedbackAcc} code evolves the joint system of evolution equations for the IMRI and DM in Eqs.~\eqref{eq:r2evolution} and~\eqref{eq:f_evolve}.
During the evolution, the DM distribution undergoes transient and lasting changes from the feedback associated with DF and SA; the transient effects are localized in radius around the instantaneous position of the secondary.
A consequence of this, which was discussed extensively in~\cite{Kavanagh:2020cfn,Nichols:2023ufs,Wade:2025rkk}, is that if we want to evolve the IMRI from an initial separation of $r_\fy$ with initial conditions that are consistent with an adiabatic inspiral from larger radii, we must start the simulation at a somewhat larger initial radius to obtain the desired initial conditions at $r_\fy$. 
As in~\cite{Kavanagh:2020cfn,Nichols:2023ufs,Wade:2025rkk}, we choose this initial radius $r_{2,\I}$ to be $3 r_\fy$.
This choice, however, will induce artifacts on the DM density at the end of the inspiral at locations with radii $\gtrsim 3 r_\fy$; these effects will be identified (and commented upon) in Sec.~\ref{sec:results}.

In cases in which SA occurs and there is no evaporation from the NS, $m_2$ will change over the course of the inspiral (and consequently so will the chirp mass).
Because we initialize $m_2$ to be $m_{2,0} = 1.4~\Msolar$ at $r_2 = 3r_\fy$, the secondary's mass will be slightly larger at $r_2 = r_\fy$.
As in~\cite{Wade:2025rkk}, we denote the secondary mass at $r_2 = r_\fy$ in the case $\alpha$ to be 
\begin{equation}
    m_\fy^{(\alpha)} = m_{2,0} + \delta m_\fy^{(\alpha)} ,
\end{equation}
where $\delta m_\fy^{(\alpha)}$ is the change in mass from the initial value at $r_2 = 3r_\fy$.
For reasons that are discussed in more detail in~\cite{Wade:2025rkk}, we do not compute dephasings between systems with different values of $m_\fy$, and instead we compare with ``equivalent'' vacuum systems.
Specifically, the appropriate vacuum system for a given case $\alpha$ was computed by first choosing it to have the mass $m_\fy^{(\alpha)}$.
The dephasing between these cases was computed from a radius $r_2$ given by the value of $r_\fy$ for the vacuum system with mass $m_2 = m_\fy^{(\alpha)}$.
This radius was denoted by $r_\fy^{(\alpha)}$ in~\cite{Wade:2025rkk}, which will also be used in this paper.

\subsection{Gravitational-wave observables}

We compute the number of gravitational-wave cycles from some reference time $t$ to the time the binary reaches the ISCO for a given case ``$\alpha$'', $t^{(\alpha)}_{\ISCO} > t$, by integrating twice the orbital frequency between these two times:
\begin{equation} \label{eq:NcyclesA}
        N_\mathrm{cycles}^{(\alpha)}(t) = \frac 1\pi \int_t^{t^{(\alpha)}_{\ISCO}} \! \Omega(t') \, dt' . 
\end{equation}
We use twice the orbital period, because we assume an adiabatic inspiral, and we compute just the leading, quadrupolar part of the GW phase.
We also assume $\Omega(t)$ is the instantaneous Keplerian orbital frequency $\Omega(t) \approx \sqrt{G m_1 / [r_2(t)]^3}$.
The gravitational-wave dephasing between two cases ``$\alpha$'' and ``$\beta$'' is just the difference in the number of cycles, and it will be denoted by
\begin{equation} \label{eq:DeltaNcyclesAB}
    \Delta N^{(\alpha - \beta)}_\mathrm{cycles}(t) = N^{(\alpha)}_\mathrm{cycles}(t) - N^{(\beta)}_\mathrm{cycles}(t) \, .
\end{equation}
We will also show the time-domain dephasing $\Delta N^{(\alpha - \beta)}_\mathrm{cycles}(t)$ re-expressed as a function of the instantaneous GW frequency $f_\GW$ by associating with each $t$ in the inspiral the corresponding $f_\GW = \Omega(t)/\pi$.
This convention will set the number of GW cycles and the dephasing in all cases to be zero at the ISCO frequency.
When we provide a single number for the dephasing in various tables, it will be at a particular reference time related to the radius $r_\fy^{(\alpha)}$.

The adaptive timestep in \textsc{HaloFeedback} and \textsc{HaloFeedbackAcc} is given in terms of multiples of the instantaneous orbital frequency $\Omega(t)$ of the IMRI.
For the region of the inspiral where $r_2 \leq r_\fy$, we use a maximum timestep of $10$ orbital periods, which corresponds to $20$ GW cycles.
In the interest of computational efficiency, we use a maximum timestep of $50$ orbital periods when $r_2 > r_\fy$.
In all parts of the inspiral, the timestep can be smaller than the maximum.
Given that the simulations cannot reliably resolve dynamics shorter than the timestep, we use the maximum timestep as our error estimate for the accuracy of the simulations.
Thus, we will quote dephasing numbers rounded to the nearest 20 GW cycles and show dephasing as a function of frequencies for values of $f_\GW$ where $\Delta N^{(\alpha - \beta)}_\mathrm{cycles}(f_\GW)$ is larger than 20 cycles.

\section{Results for the dephasing and dark-matter density}
\label{sec:results}

In this section, we discuss the results of simulations that were described in Sec.~\ref{sec:methods}.
First, in Sec.~\ref{subsec:dephase_spike}, we discuss the GW dephasing for the cases in Sec.~\ref{sec:DMmodels_spike} that allow for DM spikes.
In Sec.~\ref{subsec:dephase_plateau}, we next cover the analogous results for the cases with DM self-annihilation plateaus, which were introduced in Sec.~\ref{sec:DMmodels_plateau}.
Finally, in Sec.~\ref{subsec:density_final}, we show the DM density in both cases presented in Secs.~\ref{sec:DMmodels_spike} and~\ref{sec:DMmodels_plateau}.

\subsection{Gravitational-wave dephasing for the DM-spike cases} \label{subsec:dephase_spike}

\begin{table*}[t!]
    \centering
    \caption{\textbf{Number of gravitational-wave cycles and dephasing for inspirals with different DM models with spike profiles}. 
    As in  Eq.~\eqref{eq:NcyclesA}, the number of GW cycles for an IMRI with a primary of mass $m_1$ and a secondary is denoted $N_\mathrm{cycles}^{(\alpha)}$ with $\alpha=\mathrm{BH}$ for a black hole (second column) or~\ref{sec:spike} and~\ref{sec:spike+SA} for the NS cases covered in Secs.~\ref{sec:spike} and~\ref{sec:spike+SA} (fifth and seventh columns).
    As in Eq.~\eqref{eq:DeltaNcyclesAB}, the number of GW cycles of dephasing from vacuum IMRIs is denoted by $\Delta N_\mathrm{cycles}^\mathrm{(V-\alpha)}$ for different cases $\alpha$.
    Those shown are $\alpha=$~BH (third column) and for a NS secondary, the DM spike cases are labeled by the section headings in Sec.~\ref{sec:DMmodels_spike} (specifically, $\alpha=$~\ref{sec:spike} is without SA, \ref{sec:spike+SA} is with SA, and \ref{sec:spike+SA+no_growth} is with SA but no mass increase; they are given in the sixth, eighth, and tenth columns, respectively).
    The total cycles and dephasing correspond to the last four years of inspiral, between an initial separation $r_\fy^\mathrm{(\alpha)}$ and the ISCO.
    The change in mass $\delta m_\fy^\mathrm{(\alpha)}$ between $3 r_\fy$ and $r_\fy$ is given in the fourth and ninth columns for the $\alpha = $~BH and \ref{sec:spike+SA} cases, in which $m_2$ evolves.
    }
    \begin{tabular}{cccccccccc}
    \hline
    \hline
    & \multicolumn{3}{c}{$\overbrace{\hspace{15em}}^\text{\small BH secondary}$} & \multicolumn{6}{c}{$\overbrace{\hspace{31em}}^\text{\small NS secondary}$} \\
    $m_1 [\mathrm M_\odot]$ & $N_\mathrm{cycles}^{\mathrm{(BH)}}$ & $\Delta N_\mathrm{cycles}^\mathrm{(V-BH)}$ & $\delta m_\fy^{\mathrm{(BH)}}  [\mathrm M_\odot]$ & $N_\mathrm{cycles}^{\mathrm{(\ref{sec:spike})}}$ & $\Delta N_\mathrm{cycles}^\mathrm{(V-\ref{sec:spike})}$ & $N_\mathrm{cycles}^{\mathrm{(\ref{sec:spike+SA})}}$ & $\Delta N_\mathrm{cycles}^\mathrm{(V-\ref{sec:spike+SA})}$ & $\delta m_\fy^\mathrm{(\ref{sec:spike+SA})} [\mathrm M_\odot]$ & $\Delta N_\mathrm{cycles}^\mathrm{(V-\ref{sec:spike+SA+no_growth})}$ \\
    \hline
    $1.4\times 10^{4}$ & 2,243,320 & 12,220 & $1.76\times 10^{-3}$ & 2,244,980 & 11,620 & 2,243,140 & 12,280 & $1.95\times 10^{-3}$ & 12,160 \\
    $1.4\times 10^{3}$ & 4,011,860 & 8,920 & $6.00\times 10^{-3}$ & 4,020,560 & 6,660 & 4,011,120 & 9,100 & $6.51\times 10^{-3}$ & 8,680 \\
    \hline
    \hline
    \end{tabular}
    \label{tab:dephase_BHNS_spike}
\end{table*}

In this part, we discuss results for DM spike profiles initialized to Eq.~\eqref{eq:rhoDM}, which were obtained following the prescriptions described in Sec.~\ref{sec:methods}.
We first discuss the total dephasing from equivalent vacuum systems (see Sec.~\ref{sec:methods}), and then we discuss the dephasing as a function of GW frequency.

\subsubsection{Total gravitational-wave dephasing} \label{subsubsec:totalDephasing}

We present in Table~\ref{tab:dephase_BHNS_spike} the total number of cycles and the GW dephasing against vacuum systems for inspirals with a BH secondary or NS secondary.
We also list the $\delta m_\fy^{(\alpha)}$ for the BH case (denoted by $\alpha = \mathrm{BH}$), and the NS case assuming capture and no evaporation (denoted by $\alpha = $~\ref{sec:spike+SA}).

Because cases with different values of $\delta m_\fy^{(\alpha)}$ use different comparable vacuum systems to compute the dephasing (as well as the dephasing being computed from different $r_\fy^{(\alpha)}$), their dephasing values are not straightforward to compare (or to compare with the scenarios~\ref{sec:spike} or~\ref{sec:spike+SA+no_growth}).
Thus, we first discuss general trends in the dephasing values in Table~\ref{tab:dephase_BHNS_spike} for the different mass ratios and when including different physical effects.
A trend across all cases is that the dephasing values against vacuum are larger for a mass ratio of $q=10^{-4}$ than for $q=10^{-3}$.
This is consistent with prior work~\cite{Kavanagh:2020cfn,Nichols:2023ufs,Wade:2025rkk} (though note the different convention for computing the dephasing in~\cite{Kavanagh:2020cfn}).
The other clear trend is that including SA, and having $m_2$ increase with time both cause more dephasing against vacuum.

Because in all cases the dephasing values are between $7\times 10^3$ and $1.2\times 10^4$ (to the nearest thousand), the results in~\cite{Coogan:2021uqv} imply that all the GWs from these systems are likely to be distinguishable by LISA from IMRIs without a DM environment.
We would also like to be able to determine if BH and NS secondaries are likely to be distinguishable, and whether the different DM scenarios in Sec.~\ref{sec:DMmodels_spike} for a NS secondary could be disentangled.
We discuss how this could be assessed next.

Cases~\ref{sec:spike} and~\ref{sec:spike+SA+no_growth} have $\delta m_\fy^{(\alpha)} = 0$, so it is reasonable to take the difference of the respective dephasing values in these cases.
The difference $\Delta N_\mathrm{cycles}^\mathrm{(\ref{sec:spike}-\ref{sec:spike+SA+no_growth})}$, shows how the accretion of DM particles onto the NS, without a corresponding gain in mass, increases the rate of inspiral.
Given that this change in the number of GW cycles is hundreds ($q=10^{-4}$) or thousands ($q=10^{-3}$), the results of~\cite{Coogan:2021uqv} suggest that scenarios with and without SA could be distinguished. 

In a similar vein, we may compare the total number of cycles in the final four-year period for the different cases.
We observe that $N_\mathrm{cycles}^{\mathrm{(BH)}}$ is larger than $N_\mathrm{cycles}^{\mathrm{(\ref{sec:spike+SA})}}$ by $180$ cycles for the $m_1 = \unit[1.4 \times 10^4]{\Msolar}$ case and $740$ cycles for $m_1 = \unit[1.4 \times 10^3]{\Msolar}$.
These $O(10^{-4})$ fractional changes in the number of cycles are related to the slightly larger SA cross section for the NS, which produces more SA and speeds up the inspiral slightly.
While these dephasing numbers look promising from the perspective of detection, it should be noted that in both cases there is an order $10^{-4}$ fractional difference in the mass $m_2$ at $r_\fy$ in these two cases.
The effect of the time-dependent mass $m_2$ has the largest effect on the evolution of $r_2$ (and thence the number of cycles) from the gravitational radiation-reaction term $\dot r_2^\GW$ in Eq.~\eqref{eq:r2evolution}.
Given that the $O(10^6)$ number of cycles is determined primarily by radiation reaction, this mass difference causes a change in the chirp mass that will affect the dephasing by hundreds of cycles.
Because the secondary mass at $r_\fy$ will not be known \emph{a priori}, it would be beneficial to compare cases with the same secondary mass at $r_\fy$.

As a simple proxy for this, we can also compare (subject to the caveats above) the dephasing values in the third and eighth columns of Table~\ref{tab:dephase_BHNS_spike} to understand the possible distinguishability of the BH and~\ref{sec:spike+SA} cases.
This difference is smaller (roughly 60 GW cycles for $q=10^{-4}$ and 200 for $q=10^{-3}$).
Recall that the case~\ref{sec:spike+SA} corresponds to the limit of large DM-matter interaction cross section, such that all dark matter in the path of the neutron star is accreted.
In this limit, we noted in Sec.~\ref{sec:spike+SA} that the accretion cross section of a NS (for $m_2 = 1.4 \, \Msolar$) is about $10 \%$ larger than the corresponding BH cross section.
Thus, the dephasing values for the BH case can be estimated by decreasing the results in case~\ref{sec:spike+SA} by roughly 10\% of the difference between the dephasing values for the~\ref{sec:spike} and~\ref{sec:spike+SA} cases.
The actual results are close to this estimate.
Because these values are small, distinguishing between a BH and a NS for the \ref{sec:spike+SA}~case appears more challenging.

The difference between $N_\mathrm{cycles}^{\mathrm{(\ref{sec:spike})}}$ and $N_\mathrm{cycles}^{\mathrm{(\ref{sec:spike+SA})}}$ is about two thousand ($q=10^{-4}$) or nearly ten thousand ($q=10^{-3}$).
These numbers are larger than the related comparisons in~\cite{Nichols:2023ufs} for BH secondaries with a mass of $10~\Msolar$. 
These differences arise for several reasons: the lighter secondary undergoes more GW cycles, the secondary mass was fixed in~\cite{Nichols:2023ufs} (effectively making it like the case~\ref{sec:spike+SA+no_growth} here), the DM density did not have an angular-momentum cutoff in~\cite{Nichols:2023ufs}, and (as noted above) the NS and BH cross sections for SA differ.
However, because the masses in these two cases disagree by $\delta m_\fy^\mathrm{(\ref{sec:spike+SA})}$, this difference in cycles is not a good proxy for how distinguishable the two scenarios are likely to be.
Instead, we compare the sixth and eighth columns (at a given mass ratio), which correspond to the cases without SA (\ref{sec:spike}) and with SA (\ref{sec:spike+SA}), respectively.
These differences are in the hundreds ($q=10^{-4}$) or thousands ($q=10^{-3})$.
The mass ratio-dependence is similar to that in~\cite{Nichols:2023ufs} in that including SA (and a varying $m_2$) contributes more to the dephasing from vacuum at $q=10^{-3}$ than it does for $q=10^{-4}$.
Thus, having SA versus not having it is likely to be distinguishable.

Finally, the effect on the dephasing due to annihilation of dark matter in the neutron star can be seen by comparing the eighth and tenth columns of Table~\ref{tab:dephase_BHNS_spike} (cases~\ref{sec:spike+SA} and~\ref{sec:spike+SA+no_growth}, respectively).
The size of the dephasing is roughly 100 for $q=10^{-4}$ and a few hundred for $q=10^{-3}$.
While not quite as small as the BH and scenario~\ref{sec:spike+SA} comparison, these two cases also would be one of the more challenging to distinguish.
There is a similar trend in that an increasing secondary mass contributes more to the dephasing from vacuum at less-extreme mass ratios than it does at more-extreme ones.
The comparison between cases~\ref{sec:spike+SA} and~\ref{sec:spike+SA+no_growth} when combined with the earlier one between cases \ref{sec:spike+SA} and~\ref{sec:spike} indicates that the presence (or absence) of $\dot r_2^\mathrm{SA}$ in the evolution of $r_2$ has a larger effect on the GW phase than the time dependence of $m_2$ in the evolution equations does.

\subsubsection{Gravitational-wave dephasing as a function of frequency} \label{subsubsec:freqDephasing}

\begin{figure}
    \centering
    \includegraphics[width=\columnwidth]{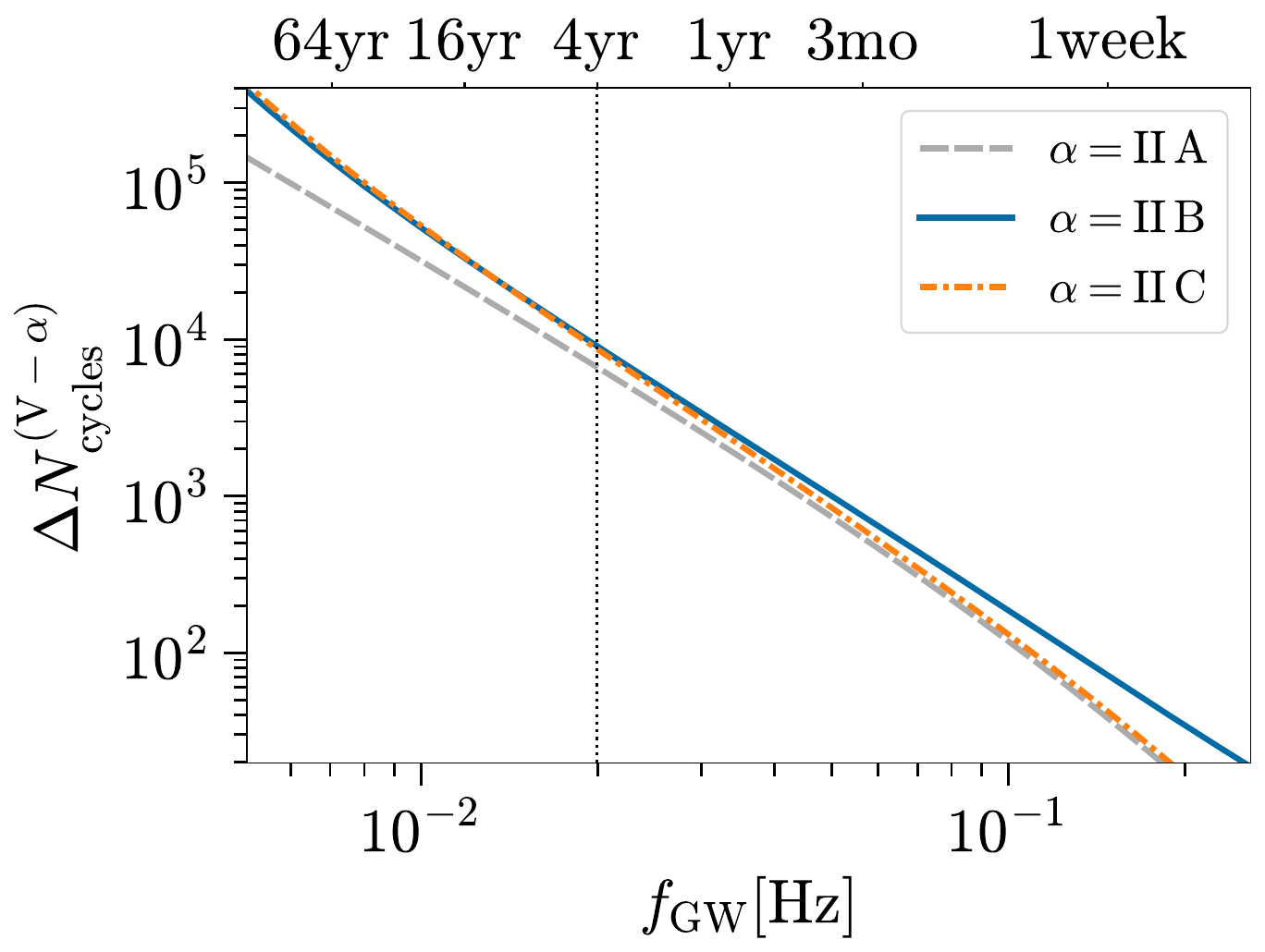} \
    \includegraphics[width=\columnwidth]{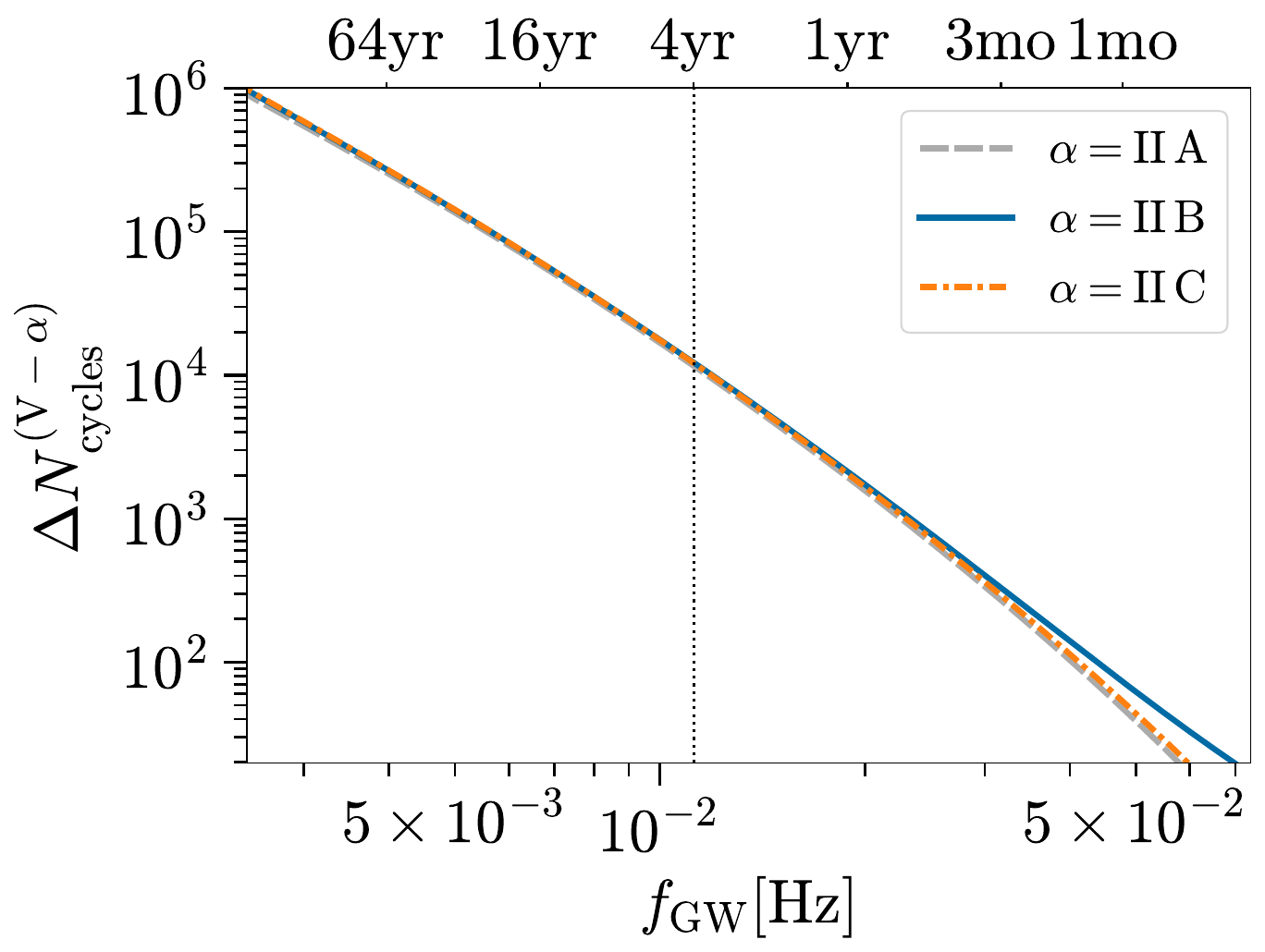}
    \caption{\textbf{Gravitational-wave dephasing for different cases in DM spike profiles}.
    We plot the gravitational-wave dephasing as a function of frequency for an IMRI with a NS secondary in a DM spike relative to a comparable vacuum IMRI.
    The dashed gray curve is the case~\ref{sec:spike} with no SA, the solid blue curve (case~\ref{sec:spike+SA}) has SA, and the dash-dotted orange curve corresponds to case~\ref{sec:spike+SA+no_growth} with evaporating capture.
    The thin vertical dotted line is the GW frequency at $r_\fy$.
    Mass ratio $q=10^{-3}$ is the top panel and $q=10^{-4}$ is the bottom panel.
    Further discussion of the figure is given in Sec.~\ref{subsec:dephase_spike}.
    }
    \label{fig:dephase_NSspikecases}
\end{figure}

It is also useful to have a more detailed view of at what stage in the IMRI's evolution the dephasing accumulates.
For this reason, in Fig.~\ref{fig:dephase_NSspikecases}, we show the dephasing against vacuum as a function of the GW frequency $f_\GW$ for the NS cases in Table~\ref{tab:dephase_BHNS_spike} in a spiked DM profile.
We show two mass ratios, where the top panel corresponds to $q = 10^{-3}$ and the bottom to $q = 10^{-4}$.
In both panels, the three curves correspond to no accretion (case \ref{sec:spike}, dashed gray), accretion and no evaporation (case \ref{sec:spike+SA}, solid blue), and accretion with evaporation (case \ref{sec:spike+SA+no_growth}, dash-dotted orange).
The frequency at $r_\fy^\mathrm{(\ref{sec:spike+SA})}$ is marked in Fig.~\ref{fig:dephase_NSspikecases} by a dotted vertical line in each panel.
At all frequencies, and in both panels, the dephasing of the \ref{sec:spike}~case has the smallest dephasing from vacuum.
In both panels, the dephasing of case~\ref{sec:spike+SA+no_growth} converges towards that of case~\ref{sec:spike} at high frequencies and case~\ref{sec:spike+SA} at low frequencies (though this is much more apparent for the $q=10^{-3}$ mass ratio).
We discuss the likely physical origin of this behavior next.

At low frequencies, cases~\ref{sec:spike+SA} and~\ref{sec:spike+SA+no_growth} are more similar because we initialize the secondary's mass to be the same at $3 r_\fy$ for all cases, but the chief difference between cases~\ref{sec:spike+SA} and~\ref{sec:spike+SA+no_growth} is that $m_2$ evolves from $m_{2,0}$ in the \ref{sec:spike+SA}~scenario and remains fixed in the \ref{sec:spike+SA+no_growth}~one.
Thus, it is reasonable that the dephasing curves will be more comparable at large radii and early times.
If we used a different prescription for initializing the masses (such as choosing them such that they agree at the ISCO), they might not agree as closely.
At low frequencies, the curves for the cases~\ref{sec:spike+SA} and~\ref{sec:spike+SA+no_growth} in the top panel have the unusual property that the dephasing against a vacuum system is larger for the \ref{sec:spike+SA+no_growth}~case than for the \ref{sec:spike+SA}~case.
This is an artifact related to the \ref{sec:spike+SA}~case being compared against a more massive vacuum IMRI (by an additive factor of $\delta m_\fy^\mathrm{(\ref{sec:spike+SA})}$) than the \ref{sec:spike+SA+no_growth}~case; this causes the vacuum number of cycles to be smaller by an amount of order $\delta m_\fy^\mathrm{(\ref{sec:spike+SA})}/m_{2,0} \sim O(10^{-3})$, which decreases the dephasing by a corresponding amount.
Given that the number of cycles approaches $O(10^7)$ at the lowest frequencies shown, this is sufficient to make the dephasing for the \ref{sec:spike+SA+no_growth}~case very slightly, but visibly, larger.

At high frequencies (smaller separations), the dephasing for the cases~\ref{sec:spike} and~\ref{sec:spike+SA+no_growth} become more similar.
This occurs because of two factors: DF effects are larger than those of SA at all frequencies (including high frequencies), and the mass $m_2$ for the two cases~\ref{sec:spike} and~\ref{sec:spike+SA+no_growth} is the same, whereas it differs for the \ref{sec:spike+SA}~case.
In all three cases, the dephasing brought about by DF is nearly the same; however, the radiation-reaction effects are the same for the cases~\ref{sec:spike} and~\ref{sec:spike+SA+no_growth} but they differ for the \ref{sec:spike+SA}~scenario.
Thus, it is reasonable that the cases~\ref{sec:spike} and~\ref{sec:spike+SA+no_growth} converge at high frequencies.

\subsection{Gravitational-wave dephasing for DM-plateau cases} \label{subsec:dephase_plateau}

We now present results for the cases with DM self-annihilation, which were discussed in Sec.~\ref{sec:DMmodels_plateau}.
In our discussion below, we again label these cases by the subsection in Sec.~\ref{sec:DMmodels_plateau} in which each scenario was introduced: namely, $\alpha = $~\ref{sec:plateau}, \ref{sec:plateau+SA}, or~\ref{sec:plateau+SA+no_growth}.
We also split the discussion into part about the total dephasing and the dephasing as a function of frequency.

\subsubsection{Total gravitational-wave dephasing}

\begin{table*}[t!]
    \centering
    \caption{\textbf{Gravitational-wave dephasing for inspirals with different DM models with DM plateaus}.
    As in Table~\ref{tab:dephase_BHNS_spike}, we list the number of gravitational wave cycles of dephasing $\Delta N_\mathrm{cycles}^\mathrm{(V-\alpha)}$ against vacuum (see Eq.~\eqref{eq:DeltaNcyclesAB}) for cases $\alpha$ of a NS secondary which now inspiral through an initial plateau density (see Sec.~\ref{sec:DMmodels_plateau}).
    We also include the total GW cycles in two cases and the change in mass $\delta m_\fy^\mathrm{(\ref{sec:plateau+SA})}$.
    Case $\alpha$=\ref{sec:plateau} corresponds to an annihilation plateau without accretion, \ref{sec:plateau+SA} to an annihilation plateau with accretion but no evaporation in the neutron star, and \ref{sec:plateau+SA+no_growth} to an annihilation plateau with accretion and evaporation in the neutron star.
    For each case, the dephasing is given for three annihilation radii (corresponding to the three values of $\rho_\mathrm{an}$) given in units of $r_\fy$, where $r_\fy = 3.71\times 10^{-8} \, \mathrm{pc}$ for $m_1=1.4\times 10^4 \, \Msolar$ and $r_\fy = 1.17\times 10^{-8} \, \mathrm{pc}$ for $m_1=1.4\times 10^3 \, \Msolar$.
    Discussion of the results in this table is given in the text of Sec.~\ref{subsec:dephase_plateau}.
    }
    \begin{tabular}{ccccccccc}
    \hline
    \hline
    $m_1 [\mathrm M_\odot]$ & $\rho_\mathrm{an} [\mathrm M_\odot/\mathrm{pc}^{3}]$ & $r_\mathrm{an} / r_\fy$ & $N_\mathrm{cycles}^{\mathrm{(\ref{sec:plateau})}}$ & $\Delta N_\mathrm{cycles}^\mathrm{(V-\ref{sec:plateau})}$ & $N_\mathrm{cycles}^{\mathrm{(\ref{sec:plateau+SA})}}$ & $\Delta N_\mathrm{cycles}^\mathrm{(V-\ref{sec:plateau+SA})}$ & $\delta m_\fy^{\mathrm{(\ref{sec:plateau+SA})}} [\mathrm M_\odot]$ & $\Delta N_\mathrm{cycles}^\mathrm{(V-\ref{sec:plateau+SA+no_growth})}$ \\
    \hline
    \multirow{3}{*}{$1.4\times 10^{4}$} & $2\times 10^{18}$ & $5.11$ & 2,256,580 & 0 & 2,255,800 & 100 & $1.16\times 10^{-3}$ & 80 \\
    & $2\times 10^{19}$ & $1.91$ & 2,256,580 & 0 & 2,255,260 & 220 & $1.84\times 10^{-3}$ & 200 \\
    & $6\times 10^{20}$ & $0.44$ & 2,244,980 & 11620 & 2,243,160 & 12280 & $1.95\times 10^{-3}$ & 12160 \\
    \hline
    \multirow{3}{*}{$1.4\times 10^{3}$} & $2\times 10^{18}$ & $7.51$ & 4,027,220 & 0 & 4,024,080 & 360 & $2.59\times 10^{-3}$ & 280 \\
    & $2\times 10^{19}$ & $2.80$ & 4,027,220 & 0 & 4,020,340 & 720 & $5.73\times 10^{-3}$ & 600 \\
    & $6\times 10^{20}$ & $0.65$ & 4,021,840 & 5380 & 4,012,460 & 7760 & $6.51\times 10^{-3}$ & 7340 \\
    \hline
    \hline
    \end{tabular}
    \label{tab:dephase_NS_plat}
\end{table*}

We begin by discussing the dephasing from vacuum IMRIs, which is summarized in Table~\ref{tab:dephase_NS_plat}.
As described in Sec.~\ref{sec:methods}, we consider three plateau densities for each mass ratio, with annihilation densities given in the second column of Table~\ref{tab:dephase_NS_plat}.
The annihilation radii corresponding to each plateau are given in the third column in units of $r_\fy$ (which is computed using $m_2 = m_{2,0}$.)
As in Table~\ref{tab:dephase_BHNS_spike}, we list the total number of cycles for the cases without and with accretion, $N_\mathrm{cycles}^{\mathrm{(\ref{sec:plateau})}}$ and $N_\mathrm{cycles}^{\mathrm{(\ref{sec:plateau+SA})}}$.
We also give the GW dephasing against comparable vacuum IMRIs for each of the DM models ($\Delta N_\mathrm{cycles}^\mathrm{(V-\ref{sec:plateau})}$, $\Delta N_\mathrm{cycles}^\mathrm{(V-\ref{sec:plateau+SA})}$, and $\Delta N_\mathrm{cycles}^\mathrm{(V-\ref{sec:plateau+SA+no_growth})}$).
We finally list $\delta m_\fy$ for the \ref{sec:plateau+SA}~case.

The columns of Table~\ref{tab:dephase_NS_plat} show that the GW dephasing values from vacuum binaries are significantly larger for the smallest annihilation radii, for which $r_\mathrm{an} < r_\fy$.
As described in Sec.~\ref{subsubsec:diffs}, DF is significantly weakened when the binary is at $r_2 < r_\mathrm{an}$.
Since the final four years of the inspiral take place with $r_2 < r_\mathrm{an}$ for the two larger values of $r_\mathrm{an}$ (smaller $\rho_\mathrm{an}$), in the case~\ref{sec:plateau} without SA, the IMRIs are indistinguishable from vacuum binaries (despite the high densities compared to those in the local region of the solar system).
For the highest density, the dephasing values are similar to the equivalent cases for the spiky DM densities (see Table~\ref{tab:dephase_BHNS_spike}), because the effects of DF are more important at larger separations, where the inspiral proceeds slowly.
There is a less substantial difference in the values of $\delta m_\fy^\mathrm{(\ref{sec:plateau+SA})}$ for different $r_\mathrm{an}$ values because most of the mass is accreted at the initial separation $r_2 \approx 3 r_\fy$ (and from Fig.~\ref{fig:density_initial}, the densities are more comparable there for different $r_\mathrm{an}$).
The values of $\delta m_\fy^\mathrm{(\ref{sec:plateau+SA})}$ are also more similar to the comparable cases for DM spikes in Table~\ref{tab:dephase_BHNS_spike}.

Because cases~\ref{sec:plateau} and~\ref{sec:plateau+SA+no_growth} have the same secondary mass, we can consider the difference $\Delta N_\mathrm{cycles}^\mathrm{(\ref{sec:plateau}-\ref{sec:plateau+SA+no_growth})}$.
When $\rho_\mathrm{an} = \unit[6\times 10^{20}]{\Msolar/pc^3}$ it is comparable to the similar dephasing for the DM spike profiles, which were discussed in Sec.~\ref{subsubsec:totalDephasing}.
For the other plateau radii, the difference $\Delta N_\mathrm{cycles}^\mathrm{(\ref{sec:plateau}-\ref{sec:plateau+SA+no_growth})}$ is smaller than the equivalent one discussed in Sec.~\ref{subsubsec:totalDephasing}, because the density is decreased from the DM spike case.
Similarly, the relatively small values of $\Delta N_\mathrm{cycles}^\mathrm{(V-\ref{sec:plateau+SA})}$ and $\Delta N_\mathrm{cycles}^\mathrm{(V-\ref{sec:plateau+SA+no_growth})}$ for these systems are consequences of the comparatively small density present at these radii as well as SA being subdominant to DF (which is strongly suppressed) in its effects on the inspiral.

Note also that if we compute the differences of the dephasing values $\Delta N_\mathrm{cycles}^\mathrm{(V-\alpha)} - \Delta N_\mathrm{cycles}^\mathrm{(V-\beta)}$, as was done in Sec.~\ref{subsubsec:totalDephasing}, the results would be qualitatively similar to the cases with DM spikes.
The main difference is that for the larger radii $r_\mathrm{an}$, the dephasing is again smaller because of the lower DM densities in these cases and significantly suppressed DF.
Thus, much of the discussions of the distinguishability in Sec.~\ref{subsubsec:totalDephasing} carries over to this part as well.

\subsubsection{Gravitational-wave dephasing as a function of frequency}

\begin{figure}
    \centering
    \includegraphics[width=\columnwidth]{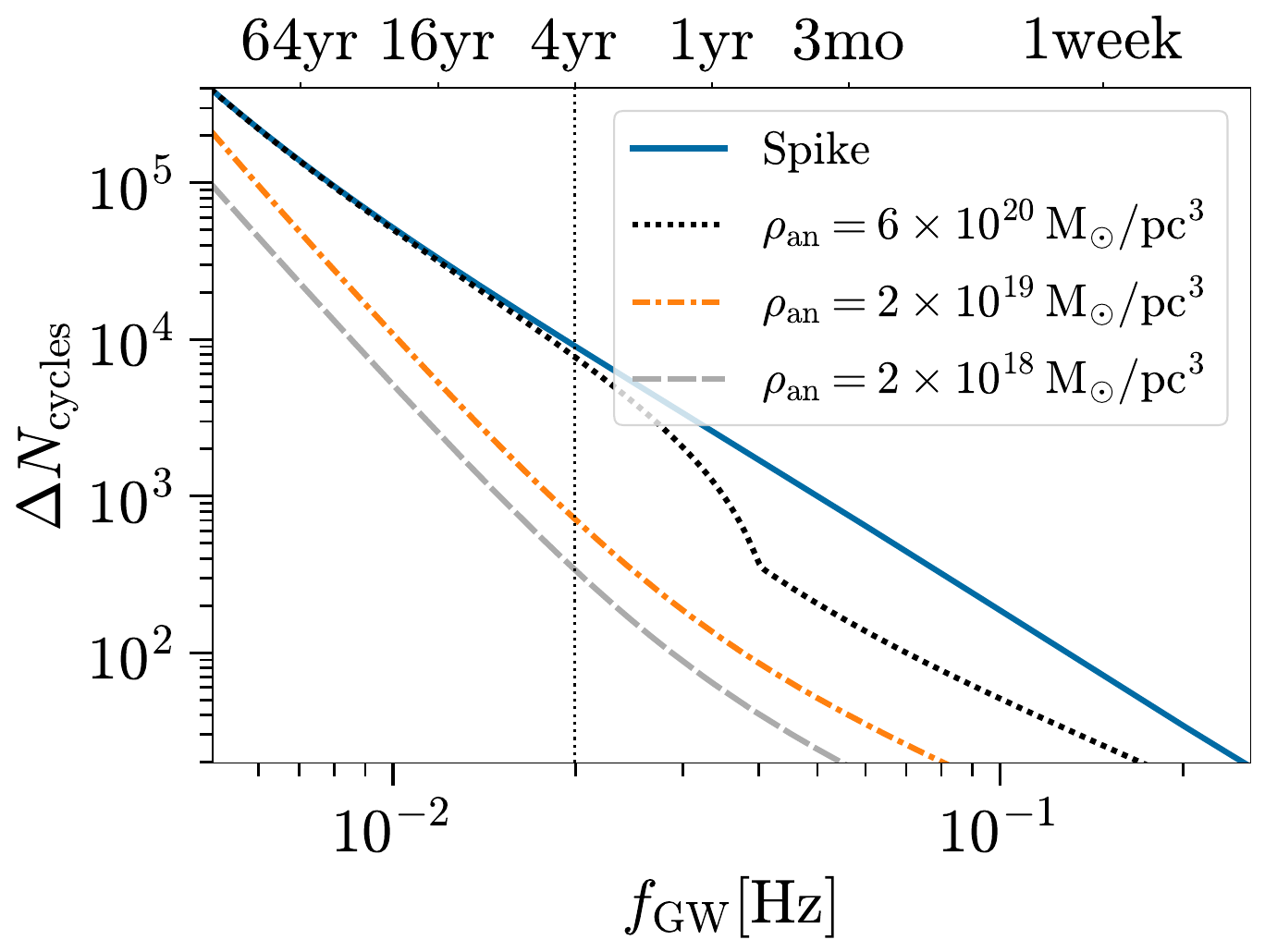} \
    \includegraphics[width=\columnwidth]{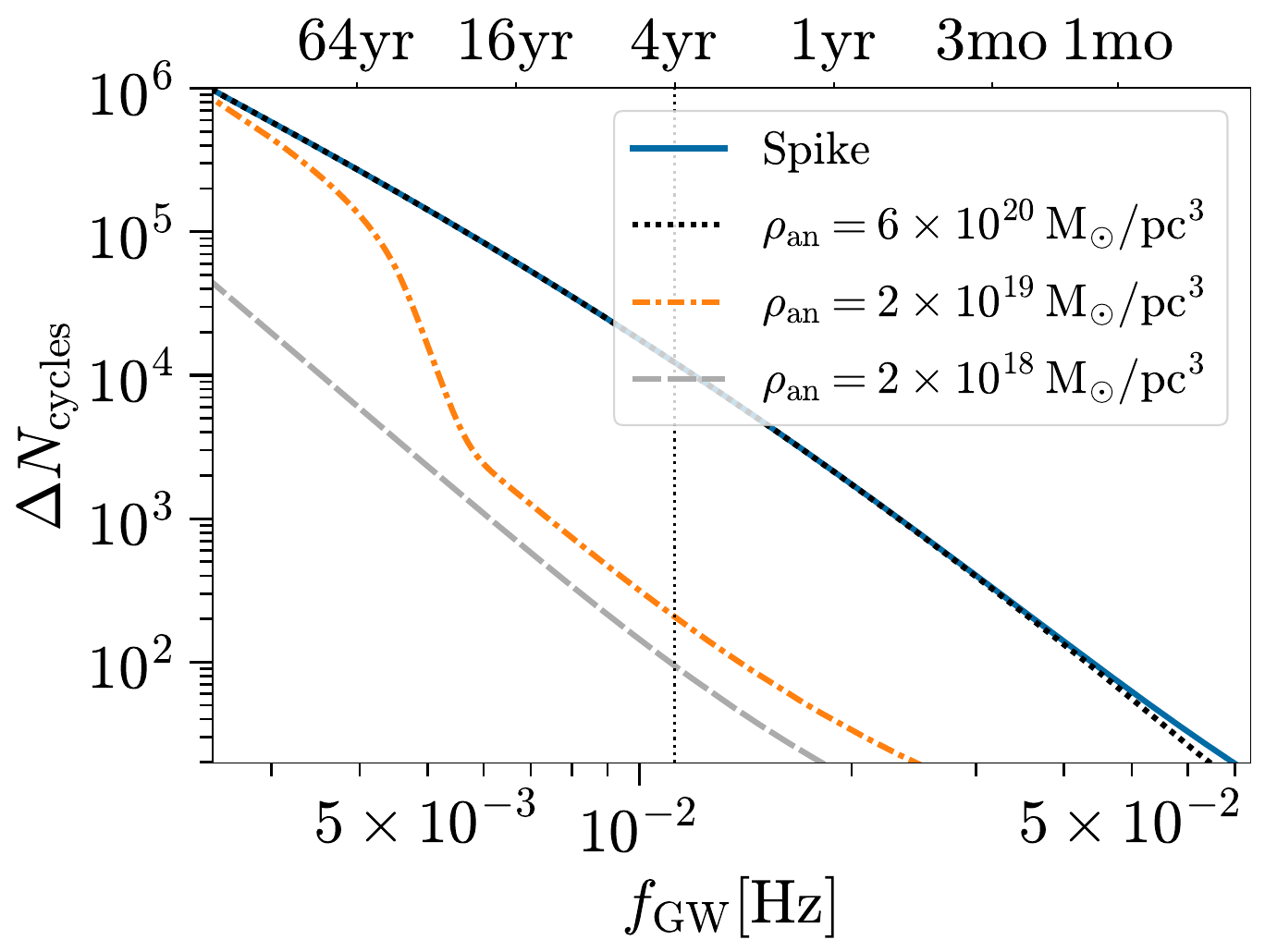}
    \caption{\textbf{Gravitational-wave dephasing for a spike profile and plateau profiles with different annihilation densities}.
    We plot the gravitational-wave dephasing of a BH-NS binary relative to a comparable vacuum system for four different initial dark-matter profiles and two mass ratios.
    The primary masses are $m_1 = 1.4 \times 10^3\,\Msolar$ ($q=10^{-3}$, \emph{top}), or $m_1 = 1.4 \times 10^4\,\Msolar$ ($q=10^{-4}$, \emph{bottom}).
    Solid blue corresponds to a DM spike (case \ref{sec:spike+SA}), while dash-dotted orange, dashed gray and dotted black correspond to annihilation plateaus (\ref{sec:plateau+SA}) with different annihilation densities $\rho_\mathrm{an}$, the values of which are given in the legend.
    The thin vertical dotted line is the GW frequency at $r_\fy$.
    Further discussion of the figure is given in Sec.~\ref{subsec:dephase_plateau}.}
    \label{fig:dephase_NScases}
\end{figure}

In Fig.~\ref{fig:dephase_NScases}, we plot the dephasing against vacuum for cases \ref{sec:spike+SA} and \ref{sec:plateau+SA} for the three values of $\rho_\mathrm{an}$.
The top panel depicts a mass ratio of $q=10^{-3}$ whereas the bottom is for $q=10^{-4}$.
The thin dotted curves again depict the frequency at $r_\fy$.
For each of the plateau profiles, the dephasing is smaller than that for the spike profile; the dephasing also decreases for larger annihilation radii, which is consistent with the decrease in density for larger annihilation radii.
The dependence of the dephasing curves as a function of frequency for the different cases in Fig.~\ref{fig:dephase_NScases} is somewhat subtle, but it will be explained next.

For context, in~\cite{Kavanagh:2020cfn}, it was noted that in a nonevolving DM spike, the GW dephasing induced by dynamical friction behaved like a $\gamma_\SP-11/2$ post-Newtonian effect (which leads to a dephasing with a power law of $(2/3)(\gamma_\SP-8)$ in the frequency domain).
Secondary accretion produces an effect at one post-Newtonian order higher (see~\cite{Nichols:2023ufs}), so its dephasing is instead $(2/3)(\gamma_\SP-7)$.
Finally, a difference in the chirp mass of two systems leads to a dephasing with a power law of $-5/3$, which is the power law of the leading GW phase.
In~\cite{Coogan:2021uqv}, it was shown that the dephasing for evolving spikes (from just DF feedback) could be modeled well by replacing $\gamma_\SP$ in the power law with $\gamma_\mathrm{eff}$, which was defined to be a power law of an effective density at the secondary's location as a function of radius.

The spike cases (solid blue curves) in the panels of Fig.~\ref{fig:dephase_NScases} behave approximately like a single power law for a $\gamma_\mathrm{eff}$ (as in~\cite{Coogan:2021uqv}), but not precisely, because we have used an angular-momentum cutoff which modifies the density from a single power law at smaller separations.
The black dotted curve in the top panel follows the spike case at low frequencies and changes between $r_\fy$ and $r_\mathrm{an}$ to a power law with a slope of approximately $-5/3$ (note that the instantaneous orbital frequency when $r_2 = r_\mathrm{an}$ is $\sqrt{G m_1 / r_\mathrm{an}^3} / \pi \approx 3.8\times 10^{-2}\, \mathrm{Hz}$).
The effects of DF decrease for radii smaller than $r_\mathrm{an}$ (larger GW frequencies), and the dephasing predominantly arises due to the fact that the secondary accreted mass between $r_\fy$ and $r_\mathrm{an}$.
This increases the chirp mass, and produces a dephasing from a comparable vacuum system where the chirp mass of the vacuum system was chosen to be equal to that of the system in a DM plateau at an orbital separation of $r_2 = r_\fy$.

In the top panel, the orange dash-dotted and gray dashed curves have a similar qualitative frequency dependence to each other (the gray dashed case is just smaller due to the lower DM density).
The annihilation radius is larger than the four-year radius in these cases (and, in both cases, the radius maps to a corresponding GW frequency that is lower than those shown in the plot).
The slope of the dephasing curves at lower frequencies are much steeper than in the spike and smaller annihilation radius.
This is consistent with the fact that the plateau has a shallower power law of $\gamma_\mathrm{an} = 1/2$.
At higher frequencies it begins to transition towards the power law of $-5/3$, similarly to the black dotted curve, but the transition is more gradual because the secondary is in the annihilation radius for all frequencies depicted in the plot.

The bottom panel of Fig.~\ref{fig:dephase_NScases} has some similarities and some important qualitative differences.
First, the spike and smallest $r_\mathrm{an}$ cases are much more similar, because as shown in the bottom panel of Fig.~\ref{fig:density_initial}, the densities are much more similar.
For the mass ratio $q=10^{-4}$, the frequency at the annihilation radius is $4.3\times 10^{-3}\, \mathrm{Hz}$, so the orange dash-dotted curve in the bottom panel behaves qualitatively more similarly to the black dotted curve in the top panel.
However, the slope of the higher-frequency portion of this curve is more consistent with a dephasing induced by SA in the equations of motion for $r_2$ rather than by the difference in the chirp mass, which has a slope of $-5/3$.
The dashed gray curve for the lower plateau density is also more consistent with the dephasing being produced primarily by SA.

\begin{figure}
    \centering
    \includegraphics[width=\columnwidth]{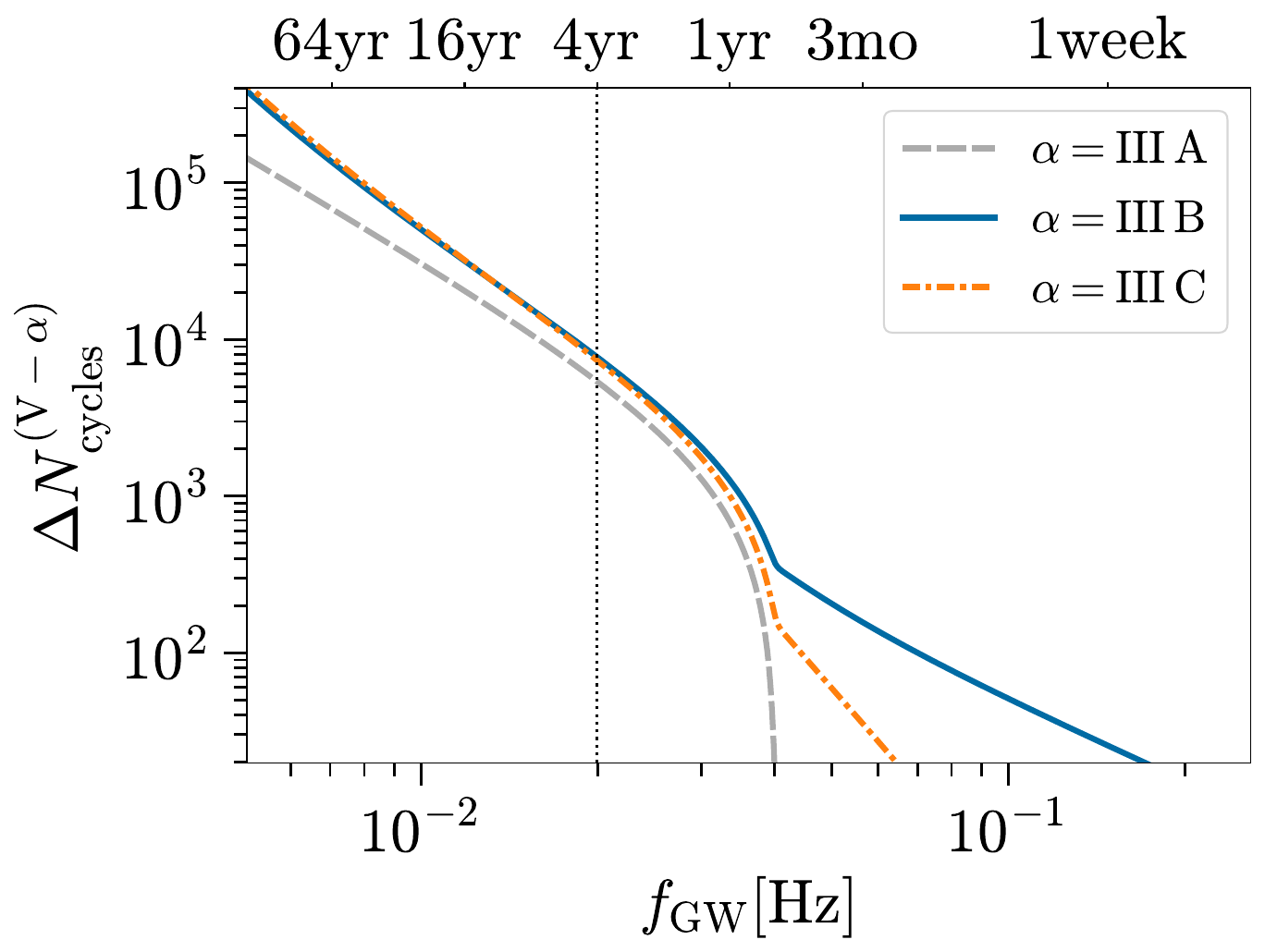} \
    \includegraphics[width=\columnwidth]{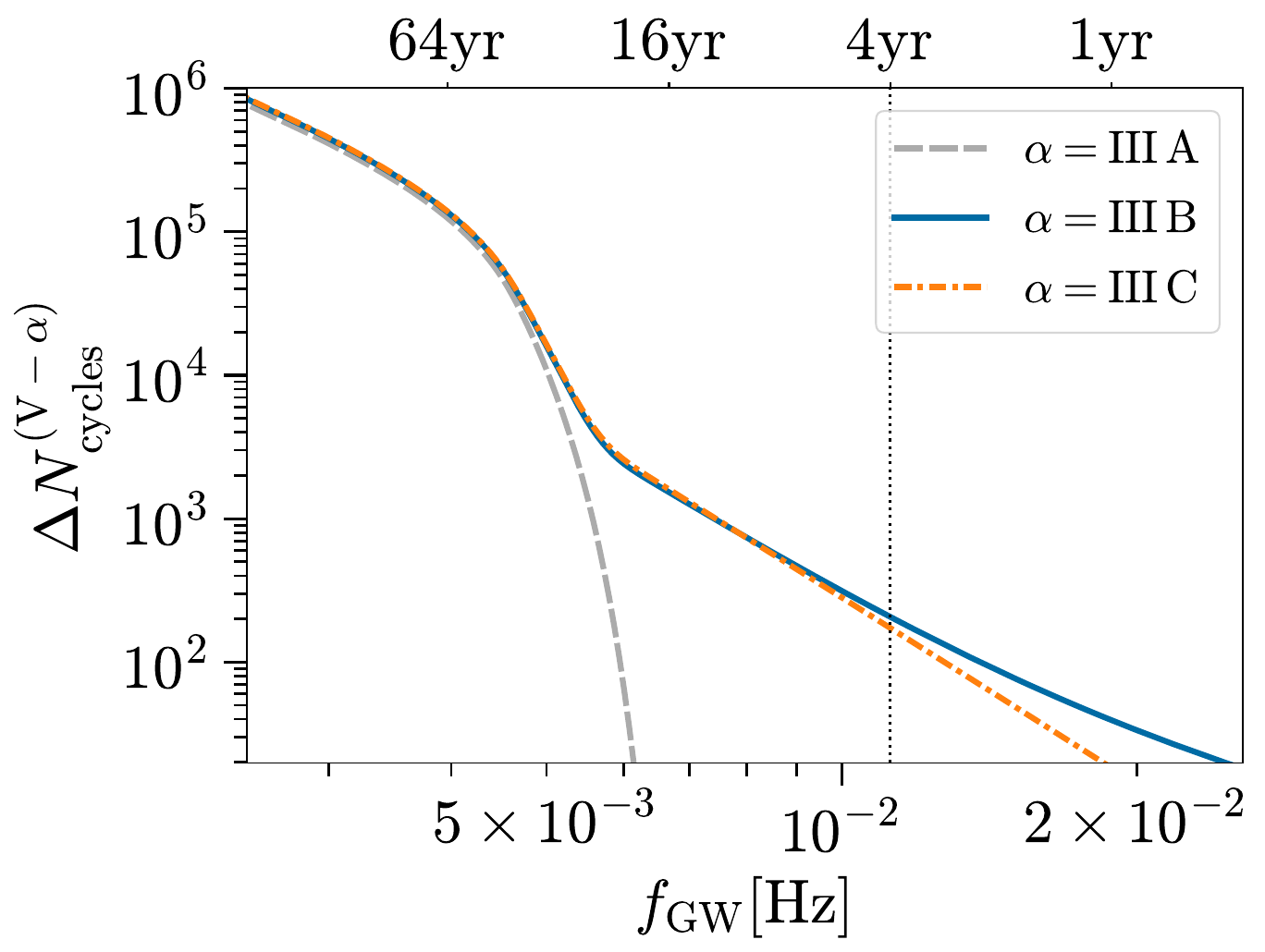}
    \caption{\textbf{Gravitational-wave dephasing for different cases in a DM plateau profile}.
    This plot is the analog of Fig.~\ref{fig:dephase_NSspikecases} for a DM plateau.
    The line styles of the curves are the same as in Fig.~\ref{fig:dephase_NSspikecases}.
    The \emph{top} panel has an annihilation density of $\rho_\mathrm{an} = \unit[6\times 10^{20}]{\Msolar/pc^3}$ and mass ratio $q=10^{-3}$; the \emph{bottom} panel has $\rho_\mathrm{an} = \unit[2\times 10^{19}]{\Msolar/pc^3}$ and $q=10^{-4}$.
    More features of the figure are discussed in the text of Sec.~\ref{subsec:dephase_plateau}.
    }
    \label{fig:dephase_NSplatcases}
\end{figure}

Next, in Fig.~\ref{fig:dephase_NSplatcases}, we plot the dephasing for three cases of no accretion (the dashed gray curve labeled \ref{sec:plateau}), accretion with no evaporation in the neutron star (the solid blue curve labeled \ref{sec:plateau+SA}), and accretion with evaporation (dash-dotted orange curve labeled \ref{sec:plateau+SA+no_growth}).
These are the analogs in a DM plateau of the DM spike cases given in Fig.~\ref{fig:dephase_NSspikecases}.
However, in Fig.~\ref{fig:dephase_NSplatcases}, we instead consider initial plateau density profiles, instead of spiked profiles.
In the top panel, we show the plateau with $\rho_\mathrm{an} = \unit[6\times 10^{20}]{\Msolar/pc^3}$ (where $r_\mathrm{an}/r_\fy = 0.65$) with $m_1 = 1.4\times 10^{3} \, \Msolar$.
The solid blue curve in the top panel of Fig.~\ref{fig:dephase_NSplatcases} is the same as the dotted black curve in the top panel of Fig.~\ref{fig:dephase_NScases}; similarly, the solid blue curve in the bottom panel of Fig.~\ref{fig:dephase_NSplatcases} is the same as the dash-dotted orange curve in the bottom panel of Fig.~\ref{fig:dephase_NScases}.
In both panels, the dephasing is smallest for the case with no accretion (\ref{sec:plateau}), next smallest for evaporating accretion (\ref{sec:plateau+SA+no_growth}) and largest for accretion without evaporation (\ref{sec:plateau+SA}).\footnote{At the lowest frequencies in the top panel, the dephasing for the \ref{sec:plateau+SA+no_growth}~case is slightly larger than the \ref{sec:plateau+SA}. 
The reason for this is the same as in the DM spike case, which was discussed in Sec.~\ref{subsubsec:freqDephasing}.}

The top and bottom panels have similar qualitative features, which we now describe.
The dashed gray curves (no SA) go to zero rapidly at radii smaller than $r_\mathrm{an}$ in each case (higher frequencies), because dynamical friction is significantly suppressed once the binary inspirals past this radius.
The case~\ref{sec:plateau+SA+no_growth} in both panels agrees with the \ref{sec:plateau+SA}~scenario at low frequencies, and approaches a different power law at higher frequencies, as DF becomes suppressed.
With DF suppressed, the only other effect of DM on the orbit in the \ref{sec:plateau+SA+no_growth}~scenario arises from the effect of SA on the evolution of $\dot r_2$; therefore, this dephasing arises from this phenomenon.  
The cases with accretion and no evaporation at the highest frequencies shown approach a power law with slope $-5/3$, as discussed above.
In the bottom panel, however, the \ref{sec:plateau+SA}~case has a region where it overlaps with the \ref{sec:plateau+SA+no_growth}~scenario before the two curves diverge at the highest frequencies depicted in the plot.

\subsection{Dark-matter density for spikes and plateaus} \label{subsec:density_final}

\begin{figure}
    \centering
    \includegraphics[width=\columnwidth]{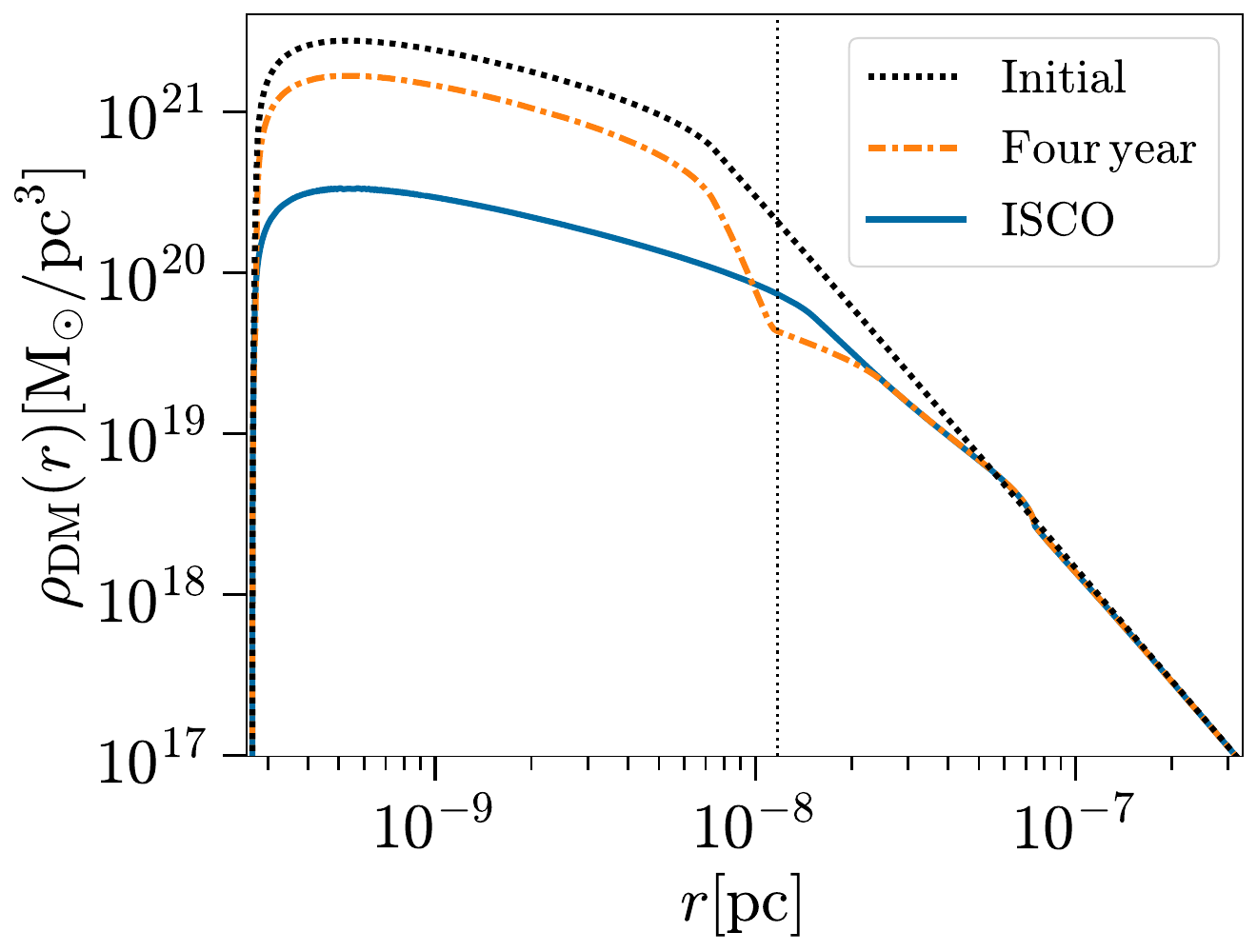} \
    \includegraphics[width=\columnwidth]{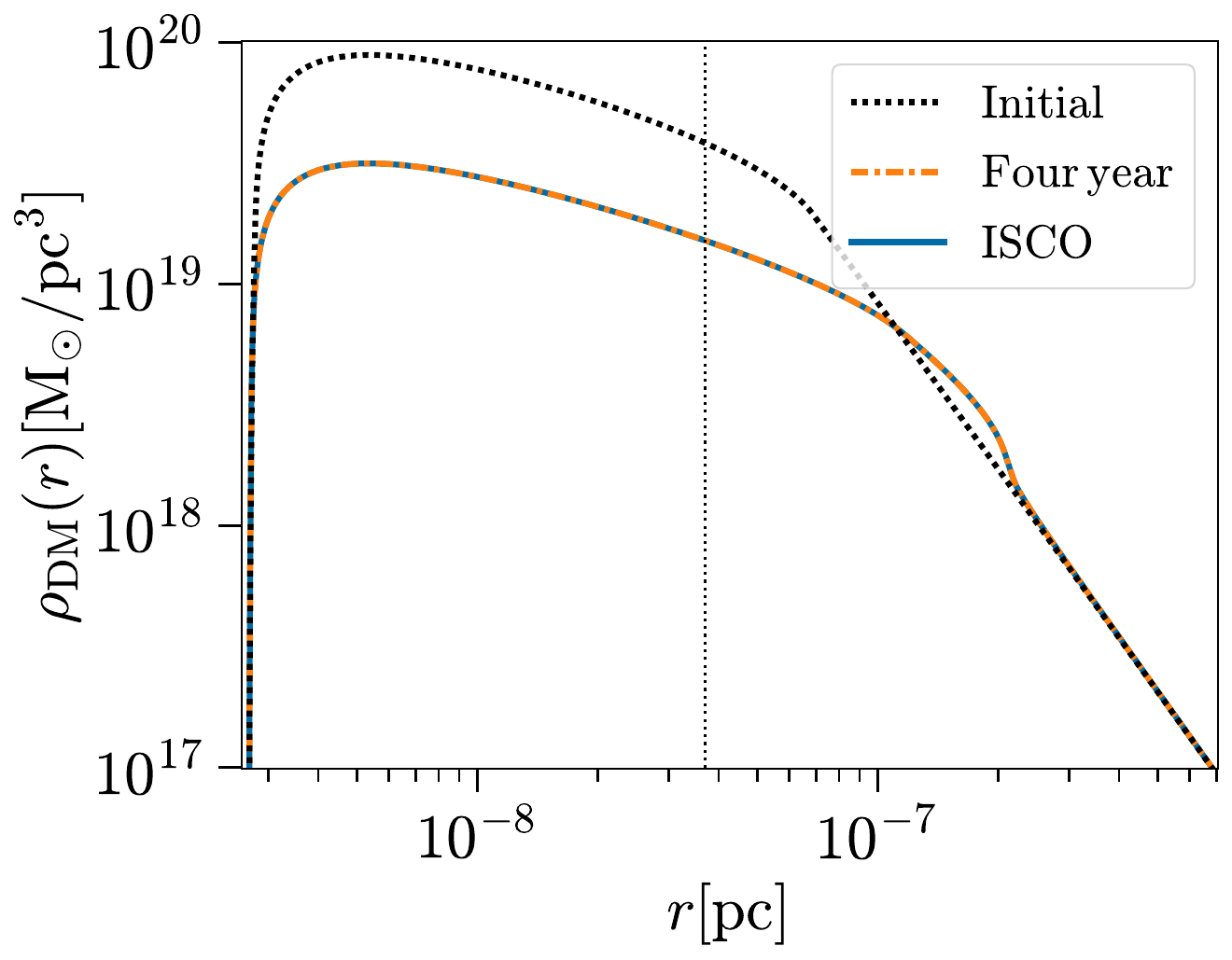}
    \caption{\textbf{Density at three stages of inspiral for self-annihilating DM with accretion onto a neutron star secondary}.
    We plot the density at multiple stages of an IMRI in a DM distribution with self-annihilation for two different primary masses.
    The dotted black curve is the initial density (where $r_2 = 3 r_\fy$), the dash-dotted orange curve is the density when $r_2 = r_\fy$, and the solid blue curve is the density at ISCO ($r_2 = r_\ISCO$).
    The thin, dotted vertical line shows the location of $r_\fy$.
    The \emph{top} panel is for a primary of $m_1 = 1.4 \times 10^3\,\Msolar$ and $\rho_\mathrm{an} = 6\times 10^{20}\,\Msolar \, \mathrm{pc}^{-3}$, corresponding to an annihilation radius of $0.65 \, r_\fy$.
    The \emph{bottom} panel is for a primary of $m_1 = 1.4 \times 10^4\,\Msolar$ and $\rho_\mathrm{an} = 2\times 10^{19}\,\Msolar \, \mathrm{pc}^{-3}$, corresponding to an annihilation radius of equivalently, $1.91 \, r_\fy$.
    The features of the figure are discussed in more detail in Sec.~\ref{subsec:density_final}.}
    \label{fig:density_NSBcases}
\end{figure}

In Fig.~\ref{fig:density_NSBcases}, we show the DM density at three stages of inspiral for the case of an inspiral through a plateau density with accretion but without evaporation (case \ref{sec:plateau+SA}).
They are the initial density, the density at a time when $r_2 = r_\fy$, and at the end of the inspiral $r_2 = r_\ISCO$.
As in Fig.~\ref{fig:dephase_NSplatcases}, the top panel is the density around a primary with $m_1 = 1.4\times 10^{3} \, \Msolar$ and with annihilation density $\rho_\mathrm{an} = \unit[6\times 10^{20}]{\Msolar/pc^3}$ (or $r_\mathrm{an} = 0.65 r_\fy$); the bottom panel has $m_1 = 1.4\times 10^{4} \, \Msolar$ and $\rho_\mathrm{an} = \unit[2\times 10^{19}]{\Msolar/pc^3}$ (or $r_\mathrm{an} = 1.91 r_\fy$).
The initial density curves are plotted in dotted black, the densities when $r_2 = r_\fy$ in dash-dotted orange, and the densities when $r_2 = r_\mathrm{ISCO}$ in solid blue.
In the bottom panel the four-year and ISCO curves overlap.

In both panels, similar features appear in the densities. 
Dynamical friction tends to scatter particles near the secondary to larger radii, which creates a pile-up of particles near the initial binary separation of $3 r_\fy$ (where $r_\fy$ is depicted using the thin, vertical dotted line).
A combination of DF and SA decreases the density at smaller radii, which causes a clear decrease in the density in the plateau and causes the radial size of the plateau region to increase.
These effects on the density are stronger for the lighter primary mass.
Additionally, for both mass ratios, the ISCO density still retains the general form of the initial profile by having a region of shallower slope at smaller radii and a region of steeper slope at larger radii.
The decrease and widening of the plateau is largely due to the secondary scattering with DM particles on eccentric orbits at larger radii (near apocenter).
The effects of DF move these DM particles onto larger radius orbits and out of the plateau, which decreases the plateau density (and increases its size).

We next highlight a few features of each panel in Fig.~\ref{fig:density_NSBcases} individually, starting with the top (where the annihilation radius is smaller than $r_\fy$).
In this case, DF is active both during the period between the initial and four-year curves and for part of the time between the four year and ISCO curves.
Because the annihilation radius is smaller than $r_\fy$, 
the effect of DF is to redistribute particles out to larger radii (as well as eject a small fraction from the spike).
The result of this is that the plateau extends out to larger radii and there is a corresponding decrease in density (which is also partly due to SA).
The slope of the curve after the inspiral is very slightly shallower than that of the initial profile ($\gamma_\mathrm{an}$).
This is consistent with the results in~\cite{Wade:2025rkk}, though not as dramatic as the results there, which assumed spike profiles with angular-momentum cutoffs rather than plateau density profiles.

In the bottom panel of Fig.~\ref{fig:density_NSBcases}, the four-year and ISCO curves very closely overlap.
Unlike the top panel, for this system the annihilation radius is larger than $r_\fy$, and consequently, DF is strongly suppressed in the region between $r_\fy$ and $r_\ISCO$.
This, combined with the environmental effects on the density being weaker for the more extreme mass ratio, explains why the four-year curve is very similar to the ISCO curve: SA alone has not significantly depleted the density in the last four years of inspiral.
This also accounts for the larger artifact of enhanced density at $3r_\fy$; in the top panel, much of this artifact is accreted because SA is more efficient at the less extreme mass ratio.

\begin{figure}
    \centering
    \includegraphics[width=\columnwidth]{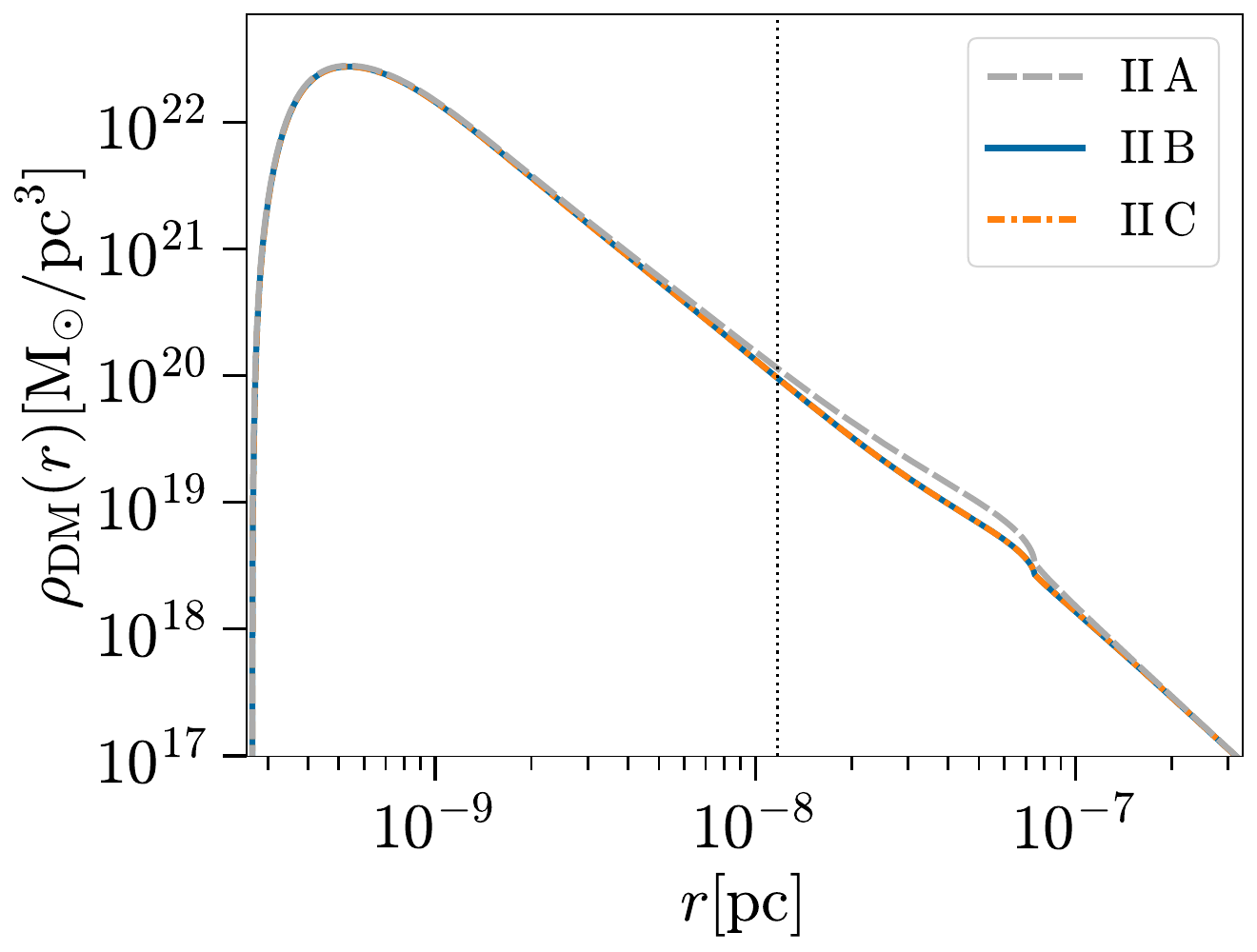} \
    \includegraphics[width=\columnwidth]{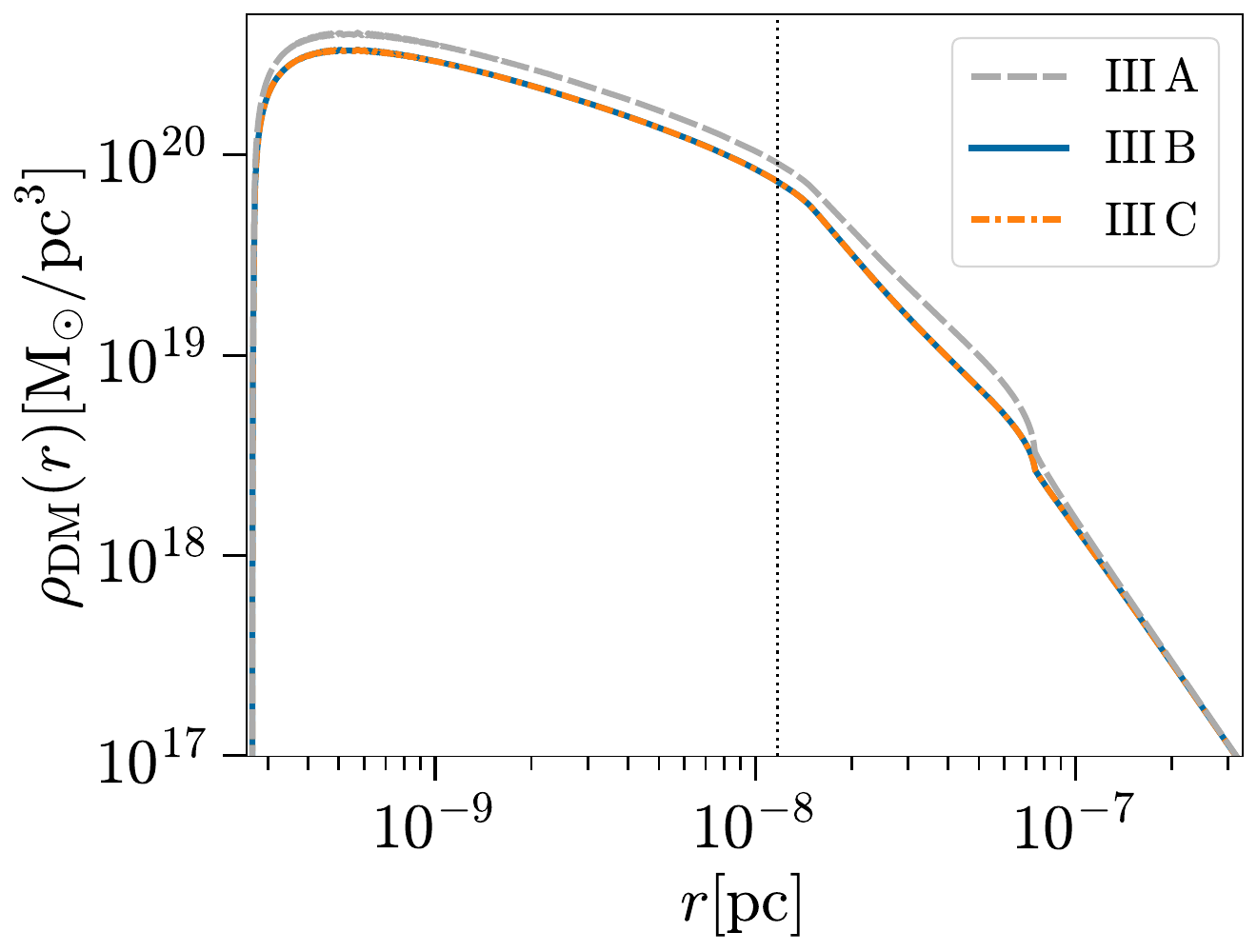}
    \caption{\textbf{Dark-matter density at ISCO for annihilating and non-annihilating dark matter}.
    We plot the density at ISCO for an IMRI in either a spike or a plateau profile, for different dark-matter models.
    In both panels, we consider systems with $m_1=1.4\times 10^{3}\, \Msolar$.
    We show the plateau profile with $\rho_\mathrm{an}=6 \times 10^{20}\,\Msolar \, \mathrm{pc}^{-3}$ ($0.65 r_\fy$) in the \emph{bottom} panel, while the spike profile is shown in the \emph{top} one.
    The color schemes for the cases in both panels are the same as those presented in Figs.~\ref{fig:dephase_NSplatcases} and \ref{fig:dephase_NSspikecases}, respectively.
    More details about the figure are discussed in Sec.~\ref{subsec:density_final}.
    }
    \label{fig:dephase_NSISCOcases}
\end{figure}

We plot in Fig.~\ref{fig:dephase_NSISCOcases} the DM density at ISCO for cases of a DM spike profile (top panel) and a plateau profile with $\rho_\mathrm{an}=6 \times 10^{20}\,\Msolar \, \mathrm{pc}^{-3}$ (bottom panel) for the primary mass $m_1=1.4\times 10^{3}\, \Msolar$.
The color scheme for the different cases is the same as that used in Figs.~\ref{fig:dephase_NSspikecases} and \ref{fig:dephase_NSplatcases}.
For both DM density profiles, the cases with and without evaporation yield final densities which are indistinguishable on the scale of the plots.
This implies that having a time-changing secondary mass does not significantly change the final density (i.e., most of the change in the density is related to the particles removed from the distribution function via SA feedback, not small changes in the rate of evolution of the IMRI).
The case without SA differs from the other two cases with it for both spikes and plateaus.
For the spike profile, the density at larger radii near $3 r_\fy$ is reduced by SA, consistent with that seen in \cite{Wade:2025rkk}.
For the plateau profile, the shape of the density is largely unchanged, but the magnitude is smaller overall.
The relatively small difference between the cases with and without SA is a strong indication that most of the decrease in density in the plateau arises from DF rather than SA.
The difference in density at small radii for the spike case is more challenging to see, given its larger magnitude as compared to the plateau case.
The spike cases are also qualitatively similar to those considered in~\cite{Wade:2025rkk}.

\section{Conclusions and discussion} \label{sec:conclusions}

In this paper, we investigated IMRI systems in which the secondary was a neutron star which inspiraled through a dense dark-matter environment around the primary, massive black hole.
We considered six different scenarios that encompassed different classes of DM models.
An important difference was whether the models allowed for DM self-annihilation or not.
Without annihilation, we assumed the DM density formed a DM spike, which smoothly truncated close to the black hole from the capture of DM particles onto the primary.
With annihilation, the innermost regions of the spike truncate to form a DM plateau, which limits the DM density to values lower than those of the DM spikes.
We then considered three different cases of interactions of NS matter with DM.
The first is purely gravitational, the second is a sufficiently efficient interaction that DM settles in the NS, and the third is an interaction that captures matter into the NS that ultimately escapes (thereby keeping the NS mass fixed).

We reviewed how these fundamental DM self-interactions and DM-NS matter interactions produce different changes in the classical equations of motion that describe the orbital evolution of the binary and the phase-space distribution of DM.
Specifically, in DM spikes, dynamical friction was present with all three classes of DM interactions in the NS.
In the first case, with only gravitational interactions, there was no effect of mass accretion onto the NS, which is present in the other two cases.
The second case also had effects on the IMRI's dynamics from the increase in mass of the NS, which are not present in the third case.
These different effects on the dynamics translate to corresponding effects on the evolution of the emitted gravitational waves (most significantly, the GW phase).
The first scenario was the most likely to be distinguishable from the other two with a GW measurement by the LISA detector, whereas the second and third scenarios would be more challenging to disentangle observationally.

For the DM plateaus, the effects on the IMRIs' orbits and the GWs depend strongly on the value of the plateau density and the radius at which it starts. 
For high densities and radii, the GW effects become similar to those of the DM spikes, whereas when the densities are sufficiently low, then the GWs become indistinguishable from vacuum systems.
The most interesting cases occurred when the IMRI's initial separation was comparable to the annihilation radius.
When it was slightly smaller, the effects of dynamical friction on the orbit become suppressed during the inspiral, which had a distinctive GW signature.
When the plateau radius was sufficiently larger than the initial radius, then there were negligible effects from DF during the inspiral, and the differences from a vacuum system arose from accretion (with or without the corresponding change in the NS mass).
Given that SA effects are weaker than those of DF, these scenarios were more challenging to distinguish from vacuum systems, despite their distinctive GW signatures.

The aims of this paper were to identify the GW signatures of these different classes of DM models and give simple quantitative calculations of their sizes.
More detailed studies of how well these different scenarios can be distinguished from one another, and exploring degeneracies in the parameter space of these DM models and different DM spike (or mound) profiles (as well as first or higher-generation mergers) is a natural avenue for future work.
Another goal for future studies would be to determine what types of constraints on the fundamental DM cross sections (for self-annihilation or interactions with NS matter) could be obtained through these GW measurements.
Doing so would require more detailed GW parameter estimation, which would itself need more efficient generation (and parameterization) of the DM effects on the emitted GWs.

\acknowledgments

The work of J.H.~was supported in part by the U.S.~Department of Energy under Grant No.~DE-SC0007974. 
D.A.N.\ and B.A.W.\ were supported in part by the NSF grant PHY-2309021 and the NSF-CAREER Award PHY-2439893.
The authors acknowledge Research Computing at The University of Virginia for providing computational resources and technical support that have contributed to the results reported within this publication.

\bibliographystyle{utcaps_mod}
\bibliography{bib}

\end{document}